\documentclass[11pt]{article}
\pdfoutput=1
\usepackage{jcappub,natbib,float,comment,bm,placeins} 
\usepackage{soul}
\usepackage[dvipsnames]{xcolor}
\usepackage{tabularx}
\usepackage{amsfonts}
\usepackage{amssymb}
\usepackage{graphicx}
\usepackage{epsfig}
\usepackage{color}
\usepackage{multirow}
\usepackage{graphicx}
\usepackage{bigints}
\usepackage{todonotes,bm,etoolbox}
\usepackage{calligra}
\usepackage{hyperref}
\usepackage{makecell}
\usepackage{booktabs}
\usepackage{subfigure}

\def\be{\begin{equation}}
\def\ee{\end{equation}}
\def\ba#1\ea{\begin{align}#1\end{align}}

\newcommand{\bq}{\begin{eqnarray}}
\newcommand{\eq}{\end{eqnarray}}

\renewcommand{\v}[1]{\mathbf{#1}}
\newcommand{\vx}{\v{x}}

\newcommand{\vk}{\v{k}}

\newcommand{\refsec}[1]{Sec.~\ref{sec:#1}}

\newcommand{\reftab}[1]{Tab.~\ref{table:#1}}

\renewcommand{\d}{\delta}

\newcommand{\bphi}{b_\phi}
\newcommand{\fnl}{f_{\rm NL}}
\newcommand{\seta}{{\rm L250N1024}}
\newcommand{\setb}{{\rm L2000N1536}}
\newcommand{\setc}{{\rm L560N1250}}

\title{Assembly bias in the local PNG halo bias and its implication for $\fnl$ constraints}

\author[a]{Titouan Lazeyras,}
\author[b,c]{Alexandre Barreira,}
\author[d]{Fabian Schmidt,}
\author[e,f]{Vincent~Desjacques}
\affiliation[a]{Dipartimento di Fisica G. Occhialini, Universit\`a degli Studi di Milano-Bicocca, Piazza della Scienza 3, 20126 Milano, Italy}
\affiliation[b]{Excellence Cluster ORIGINS, Boltzmannstraße 2, 85748 Garching, Germany}
\affiliation[c]{Ludwig-Maximilians-Universit\"at, Schellingstraße 4, 80799 M\"unchen, Germany}
\affiliation[d]{Max–Planck–Institut f\"ur Astrophysik, Karl–Schwarzschild–Stra\ss e 1, 85748 Garching, Germany}
\affiliation[e]{Physics department, Technion, 3200003 Haifa, Israel}
\affiliation[f]{Asher Space Research Institute, Technion, 3200003 Haifa, Israel}
\emailAdd{titouan.lazeyras@unimib.it, alex.barreira@origins-cluster.de, fabians@mpa-garching.mpg.de, dvince@physics.tecnion.ac.il}

\abstract{We use $N$-body simulations to study halo assembly bias (i.e., the dependence of halo clustering on properties beyond total mass) in the density and primordial non-Gaussianity (PNG) linear bias parameters $b_1$ and $b_\phi$, respectively.  We consider concentration, spin and sphericity as secondary halo properties, for which we find a clear detection of assembly bias for $b_1$ and $b_\phi$.  At fixed total mass, halo spin and sphericity impact $b_1$ and $b_\phi$ in a similar manner, roughly preserving the shape of the linear $b_\phi(b_1)$ relation satisfied by the global halo population. Halo concentration, however, drives $b_1$ and $b_\phi$ in opposite directions. This induces significant changes to the $b_\phi(b_1)$ relation, with higher concentration halos having higher amplitude of $b_\phi(b_1)$. For $z=0.5$ and $b_1 \approx 2$ in particular, the population comprising either all halos, those with the $33\%$ lowest or those with the $33\%$ highest concentrations have a PNG bias of $b_\phi \approx 3$, $b_\phi \approx -1$ and $b_\phi \approx 9$, respectively. Varying the halo concentration can make $b_\phi$ very small and even change its sign. These results have important ramifications for galaxy clustering constraints of the local PNG parameter $f_{\rm NL}$ that assume fixed forms for the $b_\phi(b_1)$ relation. We illustrate the significant impact of halo assembly bias in actual data using the BOSS DR12 galaxy power spectrum: assuming that BOSS galaxies are representative of all halos, the $33\%$ lowest or the $33\%$ highest concentration halos yields $\sigma_{f_{\rm NL}} = 44, 165, 19$, respectively. Our results suggest taking host halo concentration into account in galaxy selection strategies to maximize the signal-to-noise on $f_{\rm NL}$. They also motivate more simulation-based efforts to study the $b_\phi(b_1)$ relation of halos and galaxies.}

\begin{document}
\maketitle
\flushbottom

\section{Introduction}
\label{sec:intro}

Observational constraints on local-type primordial non-Gaussianity (PNG) are one of the most powerful ways to gain insights about the physics behind the primordial curvature fluctuations $\mathcal{R}$. This type of PNG is defined through the equation
\be
\label{eq:png}
\phi(\vx) = \phi_G(\vx) + \fnl \left[\phi_G(\vx)^2 - \left\langle \phi_G(\vx)^2 \right\rangle\right], 
\ee
where $\phi(\vx) = (3/5) \mathcal{R}(\vx)$ is the primordial gravitational potential deep in matter domination, $\phi_G$ is a Gaussian distributed random field, and $\fnl$ is a parameter that quantifies the level of non-Gaussianity of the spatial distribution of the primordial potential \cite{2001PhRvD..63f3002K}. Standard single-field models of inflation predict no local-type PNG. A detection of $\fnl \neq 0$ would thus rule out these models and point towards multifield models of inflation, in which several degrees of freedom are involved in the production of the primordial perturbations \cite{maldacena:2003, 2004JCAP...10..006C, 2011JCAP...11..038C, Tanaka:2011aj}. A nonzero value of $\fnl$ generates a primordial bispectrum (3-point correlation function) which peaks in the squeezed limit. This has been used to constrain $\fnl = -0.9 \pm 5.1\ (1\sigma)$ from measurements of the cosmic microwave background (CMB) by the Planck satellite \cite{2020A&A...641A...9P}. The next improvements over this bound are expected to come from large-scale structure data. In particular, it is a general expectation that data from upcoming galaxy surveys will allow us to constrain $\fnl$ with better than order unity precision $\sigma_{\fnl} \lesssim 1$, which will enable to test several interesting models of inflation that predict $\mathcal{O}(1)$ values for $\fnl$ \cite{2014arXiv1412.4872D, 2014arXiv1412.4671A, 2017PhRvD..95l3513D, 2019Galax...7...71B, 2021arXiv210609713S, 2022arXiv220307506F, 2022arXiv220308128A}.

The ability of large-scale structure data to take this next step in $\fnl$ constraints relies on the fact that local PNG introduces a coupling between large and small scales in the primordial fluctuations, which coherently modulates the number density of tracers of the large-scale structure such as dark matter halos or galaxies. This leaves a series of distinctive scale-dependent imprints in the clustering pattern of galaxies that can be isolated from other physical contributions to constrain $\fnl$. Concretely, in cosmologies with local PNG, the deterministic galaxy number density contrast can be linearly expanded in Fourier space as \cite{slosar/etal:2008, mcdonald:2008, giannantonio/porciani:2010, 2011JCAP...04..006B, assassi/baumann/schmidt}
\be
\label{eq:biasexp}
\d_g(\vk,z) = b_1(z) \d_m(\vk,z) + \bphi(z) \fnl\phi(\vk),
\ee
where $\vk$ is the wavenumber, and $\d_m$ is the evolved matter density contrast. The parameters $b_1$ and $\bphi$ are called {\it bias parameters} and quantify, respectively, the response of the galaxy number density to the amplitude of large-scale total matter and primordial potential perturbations (see Ref.~\cite{Desjacques:2016} for a review about galaxy bias). The primordial potential and the matter density contrast are related through $\delta_m(\vk,z) = \mathcal{M}(k,z)\phi(\vk)$. Here, $\mathcal{M}(k,z) = (2/3)k^2T_m(k)D_{\rm md}(z)/(\Omega_{m}H_0^2)$ with $D_{\rm md}(z)$ being the linear growth function normalized to the scale factor $a = (1+z)^{-1}$ during matter domination, $T_m(k)$ the matter transfer function, $\Omega_m$ the present-day mean matter density and $H_0$ the Hubble constant. On large scales, $T_m(k) \to 1$ so that the galaxy density contrast becomes
\bq\label{eq:biasexp2}
\delta_g(\vk, z) = \Bigg[b_1(z) + \frac{3\Omega_{m}H_0^2}{2k^2D_{\rm md}(z)}\bphi(z)\fnl\Bigg]\delta_m(\vk, z),
\eq
i.e., if $\fnl \neq 0$ the galaxy distribution is related to the total matter field in a scale-dependent way. This effect was first discovered in Ref.~\cite{dalal/etal:2008} where it was coined the {\it scale-dependent bias effect}. At the power spectrum (or 2-point correlation function) level, this gives rise to two leading-order contributions $\propto b_1\bphi\fnl/k^2$ and $\propto (\bphi\fnl)^2/k^4$ that modify the amplitude of the galaxy power spectrum on large scales. Using measurements of the power spectrum of quasars in the eBOSS survey, Ref.~\cite{2021arXiv210613725M} constrained $\fnl = -12 \pm 21\ (1\sigma)$ from this effect (see also Ref.~\cite{2019JCAP...09..010C}). The scale-dependent bias effect also imprints a signature in the galaxy bispectrum (see e.g.~Ref.~\cite{2022arXiv220615450C} for a recent study of the information content of the halo power spectrum and bispectrum on local $\fnl$). In combination with the galaxy power spectrum, this has been used to derive the constraints $\fnl = -30 \pm 29\ (1\sigma)$ \cite{2022arXiv220111518D} and $\fnl = -33 \pm 28\ (1\sigma)$ \cite{2022arXiv220401781C} from BOSS data.

As is apparent from Eq.~(\ref{eq:biasexp}), what the galaxy distribution primarily constrains through the scale-dependent bias effect is the product $\fnl\bphi$, and not just $\fnl$. Therefore, we need  a prior on $\bphi$ in order to measure $\fnl$. Noting that the parameter $b_1$ can be fit for using the data on small scales (e.g.~through combined power spectrum and bispectrum analyses), the standard approach in the literature is to assume a one-to-one relation between $\bphi$ and $b_1$. This fixes $\bphi$ in terms of the empirically determined $b_1$, thereby allowing to constrain the numerical value of $\fnl$. The most popular relation is that obtained from assuming universality of the halo mass function, which yields $\bphi(b_1) = 2\delta_c(b_1 - 1)$, where $\delta_c \approx 1.686$ is the critical density for spherical collapse. The bias parameters, however, are the response of galaxy formation to large-scale perturbations, and are thus functions of the complex astrophysical processes that govern galaxy formation. Therefore, they are quite uncertain and there is no compelling reason to expect simple relations like the universality one to hold with generality for all types of tracers of the large-scale structure.

Indeed, a number of works have pointed to the breakdown of the universality relation. For instance, multiple works have shown that the $\bphi(b_1)$ relation of halos in gravity-only simulations drops slightly below the universality relation for $b_1 \gtrsim 2$ \cite{grossi/etal:2009, desjacques/seljak/iliev:2009, 2010MNRAS.402..191P, baldauf/etal:2015, 2017MNRAS.468.3277B, 2020JCAP...12..013B}. More recently, Refs.~\cite{2020JCAP...12..013B, 2022JCAP...01..033B, 2022JCAP...04..057B} extended the study of the $\bphi(b_1)$ relation beyond gravity-only dynamics using new separate universe hydrodynamical simulations of galaxy formation with the IllustrisTNG model \cite{Pillepich:2017jle, 2019MNRAS.488.2079B}. These works have also shown that the universality relation is not a good description of the clustering of the simulated tracers. In addition, the $\bphi(b_1)$ relation can vary significantly across different tracers of the large-scale structure including galaxies selected by total and stellar mass \cite{2020JCAP...12..013B}, galaxies selected by color, black hole mass and black hole mass accretion rate \cite{2022JCAP...01..033B}, as well as the neutral hydrogen distribution that can be mapped by $21{\rm cm}$ line intensity mapping observations \cite{2022JCAP...04..057B}. Furthermore, it is still currently unknown how the $\bphi(b_1)$ relation depends on the parameterization of the baryon feedback in galaxy formation simulations, as is the reliability of the extrapolation from results for simulated to observed galaxy samples.

Our current knowledge of the $\bphi(b_1)$ relation of actual tracers is, therefore, still very uncertain. This poses a serious problem to local PNG constraints using galaxy data since different values of $\bphi$ have a direct impact on the $\fnl$ bounds. For example, observational constraints on $\fnl$ using quasars \cite{2021arXiv210613725M, 2019JCAP...09..010C} commonly report constraints assuming two $\bphi(b_1)$ relations: the universality one and the variant $\bphi(b_1) = 2\delta_c(b_1-1.6)$ derived in Ref.~\cite{slosar/etal:2008} as an approximation for halos having recently merged. For $b_1 \approx 2.5$ typical of eBOSS quasars, the latter relation gives $\bphi$ values that are $\approx 40\%$ smaller and thus yield $\approx 40\%$ weaker constraints. There is, however, no compelling reason to expect any of these two relations to be a faithful description of the actual relation for quasars, whose true constraining power on $\fnl$ remains therefore effectively unknown. More recently, Ref.~\cite{2022arXiv220505673B} used the power spectrum of galaxies in the $12^{\rm th}$ data release of the BOSS survey to explicitly illustrate how our uncertainty on the $\bphi(b_1)$ relation currently prevents us from being able to constrain $\fnl$ using the scale-dependent bias effect. Concretely, Ref.~\cite{2022arXiv220505673B} showed that different plausible assumptions about the $\bphi(b_1)$ relation result in error bars on $\fnl$ that can differ by more than an order of magnitude for the exact same data. The impact of galaxy bias uncertainties in $\fnl$ constraints has also been investigated in Refs.~\cite{2020JCAP...12..031B, 2021JCAP...05..015M, 2022JCAP...01..033B} with idealized mock galaxy power spectrum and bispectrum data.

Without assumptions on $\bphi$, the scale-dependent bias effect can only be used to assess the significance of detection of $\fnl \neq 0$ by constraining $\fnl\bphi$. Detecting $\fnl\bphi \neq 0$ implies $\fnl \neq 0$, and so it would still be possible to rule out single-field inflation without knowing $\fnl$. However, independent confirmation of such a detection with different tracer samples, identified using different techniques or in different surveys, would still be hampered by $\bphi$ uncertainties. For instance, if two surveys report a detection of $\fnl\bphi$, information on $b_\phi$ will then be required to help discriminate between a genuine $\fnl$ signal and possible foreground systematics on large scales. Knowledge of $\bphi$ will also be necessary to assess the consistency of a detection of local PNG if one survey reports a detection of $\fnl\bphi$ and another does not.

Even in the case when a robust detection of $\fnl\bphi$ is obtained, the numerical value of $\fnl$ is what we are ultimately interested in for informed conclusions about inflation, as well as to compare and combine with CMB constraints. Also, in case $\fnl\bphi$ is never detected, then the upper bound on $|\fnl|$ is what becomes relevant for inflation tests, but this depends on $b_\phi$. Further, even for $\fnl\bphi$ constraints, a good knowledge of $\bphi(b_1)$ is still important to aim for galaxy samples that have larger $|\bphi|$ and thus greater chances to detect $\fnl\bphi$.

This all strongly motivates additional work focused on improving our current knowledge of the $\bphi(b_1)$ relation. This is the goal of this paper, in which we focus on the {\it assembly bias} signal of dark matter halos in gravity-only simulations. Although assembly bias strictly refers to the dependence of halo bias on the mass accretion history at fixed total halo mass \cite{Sheth:2004, gao:2005}, here we will loosely use this designation to denote any dependence of halo bias on halo properties beyond total mass, as is commonly done. Halo assembly bias is interesting and important to study because, at fixed halo mass, different galaxy populations may reside in halos with distinct values of secondary properties such as concentration, environment, spin or shape, and thus inherit the halo assembly bias signal of their host halo population. The body of work on the assembly bias signal for the parameter $b_1$ is extensive, and it includes works focused on properties such as formation time, concentration, spin, shape, substructure content and mass accretion rate (see e.g.~Refs.~\cite{Sheth:2004, gao:2005, Gao:2006, Wechsler:2005, Jing:2006, Croton:2006, Angulo:2007, Dalal:2008, Faltenbacher:2009, Lacerna:2012, Sunayama:2015, Paranjape:2016, Lazeyras:2016, Salcedo:2017, Mao:2017, Chue:2018, Sato-Polito:2018, Paco_18,Lazeyras:2020suj,Contreras:2021oxg,Lazeyras:2021dar}). More recently, Refs.~\cite{Angulo:2007,Lazeyras:2021dar} studied also assembly bias in the second-order density bias parameter $b_2$, and Ref.~\cite{Lazeyras:2021dar} in the tidal bias parameter $b_{K^2}$ as well. Concerning the $\bphi$ parameter, and to the best of our knowledge, the only previous simulation-based study specifically addressing assembly bias is Ref.~\cite{Reid:2010}, who showed that halos with the same mass but different formation times can have significantly different values of $\bphi$. In fact, the authors of Ref.~\cite{Reid:2010} had already alerted the community to the significant impact this may have on $\fnl$ constraints. 

We will concentrate on the impact that halo concentration, spin and shape (sphericity) have on $\bphi$ and, therefore, complement the analysis of Ref.~\cite{Reid:2010} focused on formation time. Importantly, we will simultaneously study the signature of assembly bias in both $b_1$ and $\bphi$, and consequently, in the $\bphi(b_1)$ relation, which is what is relevant for observational constraints on $\fnl$. If for example the secondary dependence on a halo property affects $b_\phi$ and $b_1$ similarly, then the approximately linear $b_\phi(b_1)$ relation, and hence the observational constraints, will not be affected. In our results, for the first time, we find evidence of assembly bias in $\bphi$ for the three halo properties. We also find that, while halo spin and sphericity have a similar impact on $b_1$ and $\bphi$ and thus affect the $\bphi(b_1)$ relation only weakly, halo concentration affects $\bphi$ appreciably more than $b_1$ (and in opposite directions), which leads to a strong impact on the $\bphi(b_1)$ relation. In regimes where at least qualitative comparisons are possible, our results for $\bphi$ agree well with the first analysis of Ref.~\cite{Reid:2010}.

The rest of this paper is organized as follows. In \refsec{method}, we describe the simulation data and methodology we employ to measure the bias parameters $b_1$ and $\bphi$.  Our results on the assembly bias signal of the parameters $b_1$ and $\bphi$ (and their relation) are presented and discussed in \refsec{results}. In \refsec{fnl} we discuss the potential impact of our results to $\fnl$ constraints using as an illustrative case the BOSS DR12 galaxy power spectrum. Finally, we summarize and conclude in \refsec{concl}. 


\section{Numerical simulation data and methodology}
\label{sec:method}

In this section we describe the numerical simulation data (the main specifications are also summarized in \reftab{sims}) and the methodology we use to estimate the bias parameters $b_1$ and $\bphi$.

\subsection{Simulations and halo catalogues}
\label{sec:sims} 

Throughout this paper, we utilize three sets of gravity-only $N$-body simulations. 

\subsubsection*{Set $\seta$}
\label{sec:seta}

The first set is called $\seta$, and it was produced with the {\sc Gadget-2} $N$-body code \cite{Springel:2005} in a box with side $L_{\rm box} = 250{\rm Mpc}/h$ and $N_{\rm p} = 1024^3$ matter tracer particles. Contrary to the other two sets of simulations, the $\seta$ set is used for the first time in this work. The initial conditions were generated at a starting redshift of $z=99$ using the code described in Ref.~\cite{sirko:2005}. The fiducial cosmology is standard flat $\Lambda$CDM (without massive neutrinos) with cosmological parameters: $\Omega_m=0.319$, $\Omega_b =0.049$, $\Omega_\Lambda = 0.681$, $h=0.67$, $n_s=0.96$, $\sigma_8 = 0.83$. These match the cosmological parameters of the Flagship simulation of the Euclid Satellite Consortium \cite{EUCLID:2011zbd}\footnote{\href{https://sci.esa.int/web/euclid/-/59348-euclid-flagship-mock-galaxy-catalogue}{https://sci.esa.int/web/euclid/-/59348-euclid-flagship-mock-galaxy-catalogue}}. The particle mass resolution is $m_{\rm p} = 1.3\times 10^{9}\ M_{\odot}/h$, and the particle snapshot data we use in this paper were written at redshift $z=1.0,\, 1.5, \, 2.0$ and $3.0$, roughly covering the expected redshift range for Euclid galaxies. The suite is composed of simulations for $N_{\rm r} = 8$ random realizations of the initial conditions. For each realization, and with the same phases, we have ran also two additional simulations with the same cosmological parameters, except $\sigma_8$ which takes on the values $\sigma_8^{\rm Low} = 0.81$ and $\sigma_8^{\rm High} = 0.85$. We use this auxiliary {\it separate universe} simulations to estimate the values of the $\bphi$ parameter as explained in \refsec{biasest} below.

The halo catalogues were generated with the Amiga Halo Finder (AHF) code \cite{Gill:2004,Knollmann:2009}. The AHF code identifies halos in simulations using a spherical overdensity algorithm. Our halo definition assumes an enclosed overdensity that is 200 times the background matter density. In addition, we restrict ourselves to the objects defined as the main or parent halos, and discard their subhalos. Finally, to ensure reasonable convergence of the halo properties we are interested in (mass, concentration, spin and sphericity), we consider only halos that have at least 200 tracer particles within their $R_{200}$ radius. For $\seta$, this implies a minimum halo mass of ${\rm log}_{10} \left[M_{200}/(M_{\odot}/h)\right] \approx 11.41$.

\subsubsection*{Set $\setb$}
\label{sec:setb}

The simulation set called $\setb$ was also run with the  {\sc Gadget-2} $N$-body code, but for a box with $L_{\rm box} = 2000{\rm Mpc}/h$ and $N_{\rm p} = 1536^3$ tracer particles. The fiducial cosmology is standard flat $\Lambda$CDM (without massive neutrinos) with parameters: $\Omega_m=0.3$, $\Omega_b =0.0455$, $\Omega_\Lambda = 0.7$, $h=0.7$, $n_s=0.967$, $\sigma_8 = 0.85$. This yields a particle mass resolution of $m_{\rm p}= 1.8\times 10^{11} \, M_\odot/h$. The initial conditions were generated with the 2LPT code \cite{Scoccimarro:1997gr,Crocce:2006ve} at $z = 99$. The snapshot data analyzed in this paper are at redshift $z=0.0, \, 1.0$ and $2.0$. There are $N_{\rm r} = 2$ different realizations of the initial conditions. For each realization, we have two separate universe variants with $\sigma_8^{\rm Low} = 0.83$ and $\sigma_8^{\rm High} = 0.87$ for the estimation of the bias parameter $\bphi$. This simulation set has been used before in Ref.~\cite{Biagetti:2020skr} to compute $\bphi$ for the whole halo population as a function of total mass (i.e., without splitting into secondary properties).

The halo finding strategy for $\setb$ is exactly as for $\seta$. The main difference is that the lower resolution of this simulation set implies a higher minimum halo mass of ${\rm log}_{10} \left[M_{200}/(M_{\odot}/h)\right] \approx 13.55$.

\subsubsection*{Set $\setc$}
\label{sec:setc}

The third simulation suite is called $\setc$, and it was run with the {\sc AREPO} code \cite{2010MNRAS.401..791S, 2020ApJS..248...32W}, whose TreePM gravity solver is similar to that of {\sc Gadget-2}. These simulations evolve $N_{\rm p} = 1250^3$ matter tracer particles in a box of side $L_{\rm box} = 560\ {\rm Mpc}/h$, and the initial conditions were generated at $z_i = 127$ with the {\sc N-GenIC} code \citep{2015ascl.soft02003S}. The fiducial cosmology is again standard flat $\Lambda$CDM (without massive neutrinos) with parameters: $\Omega_m=0.3089$, $\Omega_b =0.0486$, $\Omega_\Lambda = 0.6911$, $h=0.6774$, $n_s=0.967$, $\sigma_8 = 0.816$; these match the cosmological parameters of the IllustrisTNG galaxy formation simulations \cite{Pillepich:2017jle}. The particle mass resolution is $m_{\rm p} = 7.7\times 10^{9}\ M_{\odot}/h$. In this paper, we utilize this suite's data at $z=0.5$, which is the snapshot closest to the mean redshift of galaxies in the BOSS survey. This suite is composed of $N_r = 5$ different realizations of the initial conditions, and likewise for the other two sets, for each realization there is a pair of separate universe simulations with $\sigma_8^{\rm Low} = 0.795$ and $\sigma_8^{\rm High} = 0.836$, which we use to estimate the value of $\bphi$. This simulation setup has been used in the past by Refs.~\cite{2020JCAP...12..013B, 2022JCAP...01..033B} to study the $\bphi$ parameter, as well as the higher-order local PNG bias parameter $b_{\phi\delta}$ associated with density and primordial potential perturbations. 

For this set, the halos were identified using the friends-of-friends (FoF) algorithm that runs on-the-fly with the {\sc Arepo} code with a linking length of $b=0.2$ times the mean interparticle distance. For the FoF catalogues, and differently than for the other two SO catalogue sets, the mass of the halo is defined as the total mass inside the radius within which the overdensity is 200 times the critical background density (not the matter density). In this case we consider the objects as found by the FoF algorithm, which comprise both the main halo and subhalos; we have checked that our results from the SO and FoF catalogues are in good agreement and that any small differences have no significant impact in our main conclusions. We will utilize this halo catalogue at $z=0.5$ to roughly estimate the impact of host halo concentration on the $\bphi(b_1)$ relation of BOSS-like galaxies, which are thought to reside in halos with $\gtrsim 10^{13}\ M_{\odot}/h$. This corresponds to objects with over 1000 particles at this set's resolution, which should thus provide satisfactorily converged mass and concentration estimates.

\begin{table*}
\centering
\begin{tabular}{lcccccccccccccc}
\toprule
& Set & $L_{\rm box}$ & $N_p$ & $N_{\rm r}$  & $m_{\rm p}$ & $\Omega_m$ & $\Omega_b$ & $h$ & $n_s$ & $\sigma_8$  \\
\midrule
& $\seta$ & $250$ & $1024^3$ & $8$ & $1.3\times 10^{9}$ & $0.319$ & $0.049$ & $0.67$ & $0.96$ & $0.83$ \\
\midrule
& $\setb$ & $2000$ & $1536^3$ & $2$ & $1.8\times 10^{11}$ & $0.3$ & $0.0455$ & $0.7$ & $0.967$ & $0.85$ \\
\midrule
& $\setc$ & $560$ & $1250^3$ & $5$ & $7.7\times 10^{9}$ & $0.3089$ & $0.0486$ & $0.6774$ & $0.967$ & $0.816$ \\
\bottomrule
\end{tabular}
\caption{Summary of the simulation specifications used in this paper. The box size $L_{\rm box}$ is in units of ${\rm Mpc}/h$ and the particle mass $m_{\rm p}$ in units of $M_{\odot}/h$. For each set, there are also additional simulations with different values of $\sigma_8$ to measure the bias parameter $\bphi$ (cf.~Sec.~\ref{sec:method}).}
\label{table:sims}
\end{table*}


\subsection{Bias parameters estimation}
\label{sec:biasest}

For a given population of halos (selected in bins of total mass and/or any other secondary property), we can estimate the linear bias parameter $b_1$ using the large-scale limit of the ratio of the halo-matter cross-power spectrum $P_{\rm hm}$ and matter auto power spectrum $P_{\rm mm}$ as
\be\label{eq:b1}
b_1(z) = \lim_{k\rightarrow0} \frac{P_{\rm hm}(k,z)}{P_{\rm mm}(k,z)}.
\ee
This expression is strictly valid only deep in the linear regime of structure formation where the ratio asymptotes to a constant. To account for corrections to this, rather than fitting a constant to the ratio, we fit instead the second order polynomial $b_1 + Ak^2$, where the $Ak^2$ serves to marginalize out the leading-order nonlinear and higher-derivative  corrections (see e.g.~Ref.~\cite{Lazeyras:2017} for a justification of this). We fit the ratio in Eq.~(\ref{eq:b1}) up to $k_{\rm max} = 0.15\ h/{\rm Mpc}$. We have made convergence tests by varying $k_{\rm max}$, as well as by including or not the quadratic term in the fit to confirm the robustness of our choice of $k_{\rm max}$. For any given simulation set, our quoted values and error bars for $b_1$ correspond to the mean and standard deviation of all $N_{\rm r}$ realizations of the initial conditions.

We estimate the local PNG bias parameter $\bphi$ from the response of the halo number density to the change in $\fnl\phi$, or equivalently through the peak-background split (PBS) argument \cite{dalal/etal:2008, slosar/etal:2008}, as the logarithmic derivative of the halo number density $n_{\rm h}$ w.r.t.~the amplitude of the primordial scalar power spectrum $\mathcal{A}_s$ (or alternatively to the parameter $\sigma_8$) as
\be\label{eq:bphi}
\bphi(z) = \frac{\partial\ln n_{\rm h}}{\partial(\fnl \phi)} \equiv 4 \frac{\partial\ln n_h}{\partial\delta_{\mathcal{A}_s}} = 2 \frac{\partial\ln n_h}{\partial\delta_{\sigma_8}},
\ee
where $\delta_{\mathcal{A}_s}$ and $\delta_{\sigma_8}$ are defined as $\mathcal{A}_s = \mathcal{A}_s^{\rm Fiducial}\left(1 + \delta_{\mathcal{A}_s}\right)$ and $\sigma_8 = \sigma_8^{\rm Fiducial}\left(1 + \delta_{\sigma_8}\right)$, respectively. In cosmologies with local PNG, the primordial squeezed bispectrum effectively results in a modulation of the amplitude of the local scalar power spectrum by long-wavelength modes. That is, inside large-scale $\fnl\phi$ perturbations, structure formation effectively takes place as if in a {\it separate universe} with a different value of $\mathcal{A}_s$, which is how Eq.~(\ref{eq:bphi}) can be derived (see e.g.~Sec.~7 of Ref.~\cite{Desjacques:2016}). Concretely, we use the separate universe simulations with different values of $\sigma_8$ described above to evaluate this derivative using finite-differences as
\bq\label{eq:bphinum}
\bphi(z) = \frac{1}{|\delta_{\sigma_8}|}\left[\frac{n_{\rm h}^{\rm High}(z) - n_{\rm h}^{\rm Low}(z)}{n_{\rm h}^{\rm Fiducial}(z)} \right],
\eq
where $n_{\rm h}^{\rm Low}$, $n_{\rm h}^{\rm Fiducial}$ and $n_{\rm h}^{\rm High}$ are the number density of halos in the corresponding simulation at redshift $z$ in some bin of total mass and/or any other secondary halo property. Similarly to $b_1$, the values and error bars we quote for $\bphi$ in this paper correspond to the mean and standard deviation across all $N_{\rm r}$ realizations of the initial conditions.


\section{Halo assembly bias results}
\label{sec:results}

We now turn to the presentation and discussion of our results for $b_1$ and $\bphi$, which we will show for halo populations split into three tertiles of concentration $c$, spin parameter $\lambda$ and sphericity $s$ in a given total halo mass bin. The halo concentration is quantified using the Navarro-Frenk-White (NFW) \cite{Navarro:1996} concentration parameter $c$ measured as in Ref.~\cite{Prada:2011}. Specifically, we utilize the ratio between $V_{\rm max}$, the maximum of the circular velocity, and $V_{200}$, the circular velocity at the radius $R_{200}$, to solve for the NFW concentration $c$ using the equation
\be
\frac{V_{\rm max}}{V_{200}}=\left(\frac{0.216 \, c}{f(c)}\right)^{1/2},
\label{eq:cprada}
\ee
where $f(c) = \ln (1+c)-{c}/(1+c)$. For the halo spin, we use the spin parameter defined in Ref.~\cite{Bullock:2000}
\be
\lambda = \frac{|\v{J}|}{\sqrt{2}MVR_{200}},
\label{eq:spinparam}
\ee
where the angular momentum $\v{J}$, the mass $M$ and the circular velocity $V$ are all evaluated at $R_{200}$. Concerning the halo sphericity, we follow the works of Refs.~\cite{Faltenbacher:2009, Lazeyras:2016, Lazeyras:2020suj,Lazeyras:2021dar} and define the sphericity $s$ of a halo as
\be
s=\frac{r_3}{r_1},
\label{eq:shape}
\ee
where $r_1 \geq r_2 \geq r_3$ are the three axes of the moment-of-inertia tensor of the halo particles; $s=1$ corresponds to perfectly spherical and $s \ll 1$ to very elongated halos.

Finally, for the concentration $c$ and spin $\lambda$ rather than quoting the results in terms of their actual values in a given mass bin, we follow Refs.~\cite{Wechsler:2005, Lazeyras:2016,Lazeyras:2020suj,Lazeyras:2021dar} and choose to present our results in terms of the following reparametrized variables
\be
p_{h}'= \frac{\ln(p_h/\bar{p}_h)}{\sigma(\ln p_h)},
\label{eq:prime}
\ee
where $p_h = \{c, \lambda\}$ is the value of concentration or spin in a given tertile and mass bin, and $\bar{p}_h$ and $\sigma_{p_h}$ are the mean and standard deviation of $p_h$ over all tertiles in the same mass bin. These two properties are known to be approximately lognormal distributed at fixed mass \cite{Bullock:2000,Bullock:1999he}, and so using these reparametrized variables allows us to remove most of the mass dependence on their values and eases the plotting of the results. On the other hand, since sphericity $s$ is not clearly lognormal distributed and does not depend significantly on halo mass, we skip defining a new variable for it. 


\subsection{Assembly bias in $b_1$ and $\bphi$}
\label{sec:assembly}

\begin{figure}[t]
\centering
\includegraphics[width=\textwidth]{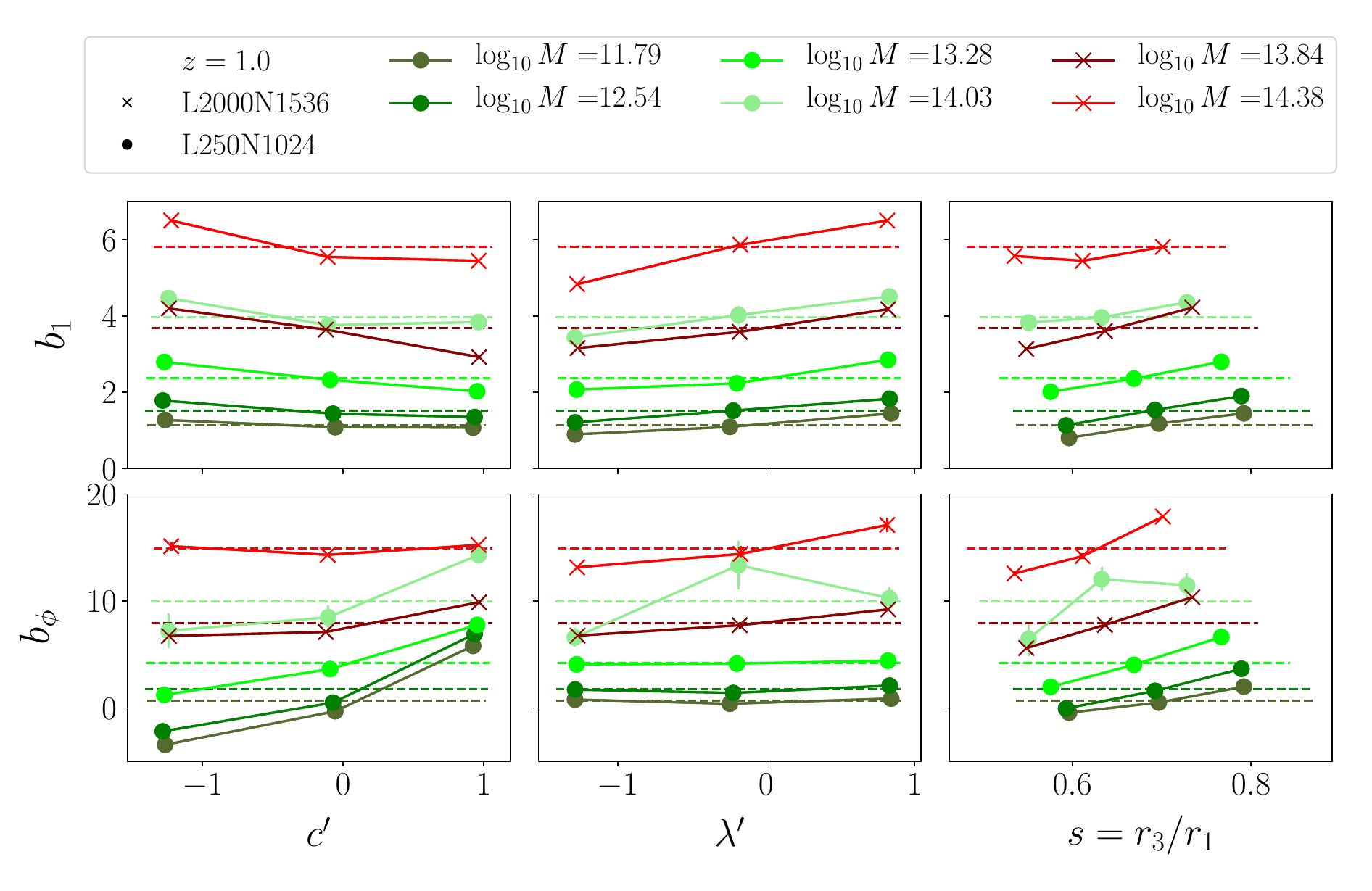}
\caption{The halo assembly bias signal in $b_1$ (top) and $\bphi$ (bottom) for concentration (left), spin (middle) and sphericity (right) as secondary halo properties. In each panel, from left to right, halos have increasing concentration, spin and sphericity. The result is shown at redshift $z=1$, and for halos selected in different total mass bins, as labeled (the ${\rm log}_{10}M$ values implicitly assume $M_{\odot}/h$ units). The two sets of symbols (circles and crosses) show the result for the $\seta$ and $\setb$ sets, respectively (note we also use two color tones to distinguish between the two simulation sets). Each set of curves shows the result for three tertiles of concentration, spin and sphericity, and the dashed lines mark the value of the bias parameter for all halos in the mass bin. The $b_1$ results are consistent with previous known results. The detection of assembly bias in $\bphi$ for these three halo properties is new to this paper (see however Ref.~\cite{Reid:2010} for a previous detection of the dependence of $\bphi$ on the halo formation time, which is in line with our concentration results).}
\label{fig:bofp}
\end{figure} 

Figure \ref{fig:bofp} shows the halo assembly bias signal in $b_1$ and $\bphi$ for concentration, spin and sphericity as secondary halo properties. In each panel, the sets of symbols connected with lines show the result for the three tertiles of the secondary halo properties in each halo mass bin, as labeled; halo concentration, spin and sphericity increase from left to right. The result is shown for the $\seta$ and $\setb$ simulation sets at $z=1$. The main takeaway points are common to other epochs, and so we display only the $z=1$ results for brevity; some of the behavior at other $z$ values can be read off from the $\bphi(b_1)$ relation in Fig.~\ref{fig:bphiofb1ofp_1} below.

For $b_1$, we recover previous known results obtained in Refs.~\cite{Lazeyras:2016, Lazeyras:2021dar}, to which we refer the reader for a more indepth discussion. Here, we focus instead on $\bphi$, whose results are new to this paper, and for which we observe a detection of assembly bias for all three secondary halo properties. Generically, the values of $\bphi$ increase with increasing values of these halo properties at fixed halo mass, but the magnitude of the effect is mass dependent. For the concentration the effect is stronger at lower masses, for the spin the effect appears stronger at higher masses, and for the sphericity the magnitude of the effect is approximately the same for all mass values shown.

The results in Fig.~\ref{fig:bofp} show further that while the assembly bias signal in $\bphi$ and $b_1$ is similar for halo spin and sphericity, it is consistently opposite for halo concentration. Specifically, for the halo mass values covered by our simulations, increasing concentration typically lowers the value of $b_1$, but typically increases the values of $\bphi$ (it becomes more positive/less negative). The absolute impact of the three secondary properties in $\bphi$ is also generally stronger than in $b_1$. It is also interesting to note that for the lowest mass values shown ${\rm log}_{10}\left[M/(M_{\odot}/h)\right] = 11.79, 12.54$, the halos in the lowest-concentration tertile have negative values of $\bphi$. This can be understood as follows. Inside positive large-scale $\fnl\phi$ perturbations (or, equivalently, given an increase in $\mathcal{A}_s$), structure formation is enhanced, and as a result, halos become generically more massive, because of enhanced mass accretion rate \cite{2020JCAP...12..013B}, and more concentrated because they form earlier. The negative values of $\bphi$ therefore indicate that, for those halo masses, the halo concentration-mass relation is enhanced by the $\fnl\phi$ perturbations, i.e., the average increase in concentration outweighs the increase in total mass of individual objects. In other words, \textit{for those masses, the number of objects with low concentration inside $\fnl\phi >0$ perturbations is lower than the cosmic mean}, hence $\bphi < 0$. We note that a more detailed study of the structure formation physics behind these results is interesting, and we defer it to future work.

It is interesting to link our halo concentration results for $\bphi$ with the previous detection in simulations by Ref.~\cite{Reid:2010} of a strong dependence of $\bphi$ on the formation time $z_f$ of halos. There, the authors find that halos that have formed earlier (old halos) have significantly larger values of $\bphi$ than halos with the same mass that have formed more recently (younger halos); see their Fig.~3, and note that in their notation, $A_{\rm NG} = \bphi/2$. We do not specifically look at the halo formation time $z_f$ and so cannot perform a detailed quantitative comparison. However, noting that halo concentration correlates with halo formation time \cite{Wechsler:2001}, with higher concentration corresponding to older halos, we find qualitative agreement between the results. Indeed, our results do show that $\bphi$ is larger for objects with higher concentration, which are normally objects that have formed a longer time ago. Reference \cite{Reid:2010} reports also that $\bphi$ can even change sign for the younger halo population, which is also in agreement with our $\bphi < 0$ results for the lowest concentration objects.


\subsection{The impact of assembly bias on the $\bphi(b_1)$ relation}
\label{sec:impact}

\begin{figure}[ht]
\centering
\includegraphics[scale=0.50]{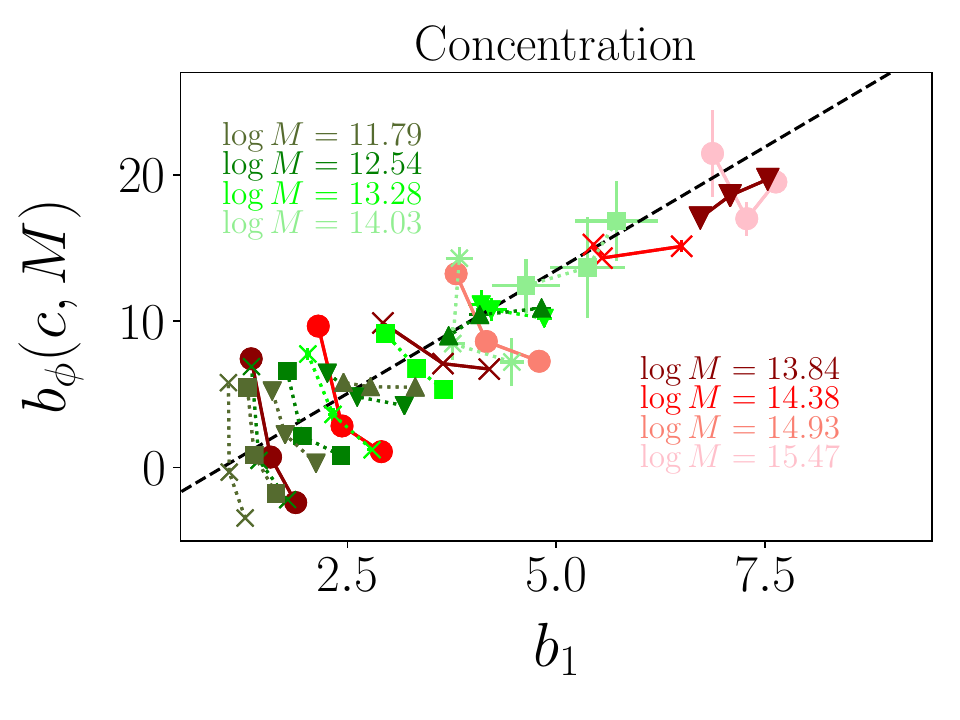}
\includegraphics[scale=0.50]{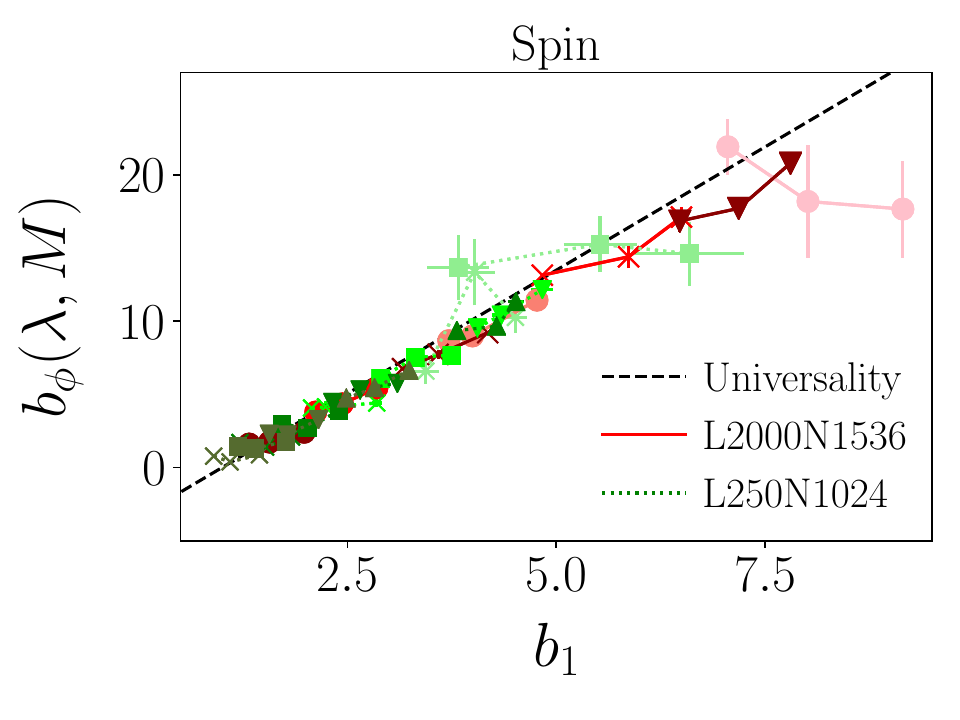}
\includegraphics[scale=0.50]{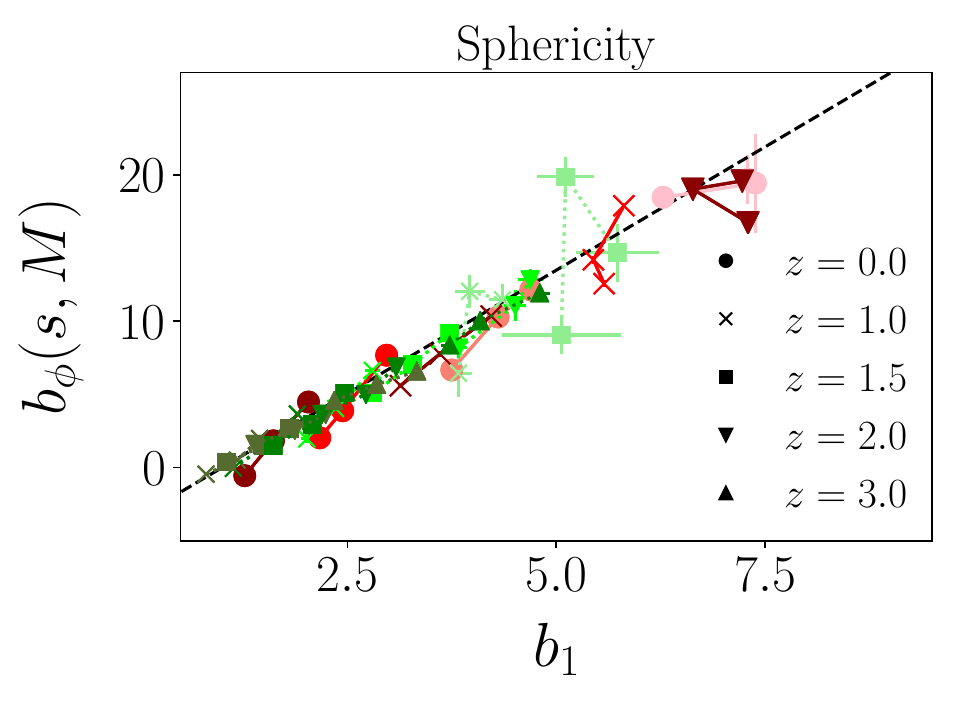}
\caption{The impact of halo assembly bias on the $\bphi(b_1)$ relation. In each panel, the sets of symbols connected with lines show the $\bphi(b_1)$ relation for halo populations with the same mass, but different concentration, spin and sphericity. The result is shown for the $\seta$ and $\setb$ simulation sets (line styles and color tone, as labeled). The color coding indicates also the halo mass values (the ${\rm log}_{10}M$ values implicitly assume $M_{\odot}/h$ units), and the different symbols indicate the redshift $z$. The universality relation $\bphi(b_1) = 2\delta_c(b_1-1)$ is shown as the dashed black line. The main takeaway is that while halo assembly bias due to spin and sphericity has a milder effect on the $\bphi(b_1)$ relation, halo concentration on the other hand has a much stronger impact (halo concentration generically decreases from top to bottom in that panel).}
\label{fig:bphiofb1ofp_1}
\end{figure}

The impact of assembly bias on the $\bphi(b_1)$ relation is shown in Fig.~\ref{fig:bphiofb1ofp_1}. As in Fig.~\ref{fig:bofp}, the sets of symbols connected by lines show the $\bphi$ and $b_1$ values for halos in the same mass bin (color coded) in different tertiles in concentration, spin and sphericity. The result is shown for the simulation sets $\seta$ (dotted lines) and $\setb$ (solid lines), and for all redshift values available for them between $z=0$ and $z=3$, as labeled. Focusing on the most observationally interesting values for $b_1 \lesssim 3$, the main takeaway is that, at fixed halo mass, spin and sphericity have no significant impact on the $\bphi(b_1)$ relation of the halos, but concentration modifies it substantially. This could have already been anticipated from Fig.~\ref{fig:bofp} for $z=1$, but it is now visible also for other redshifts. 

\begin{figure}[ht]
\centering
\includegraphics[scale=0.49]{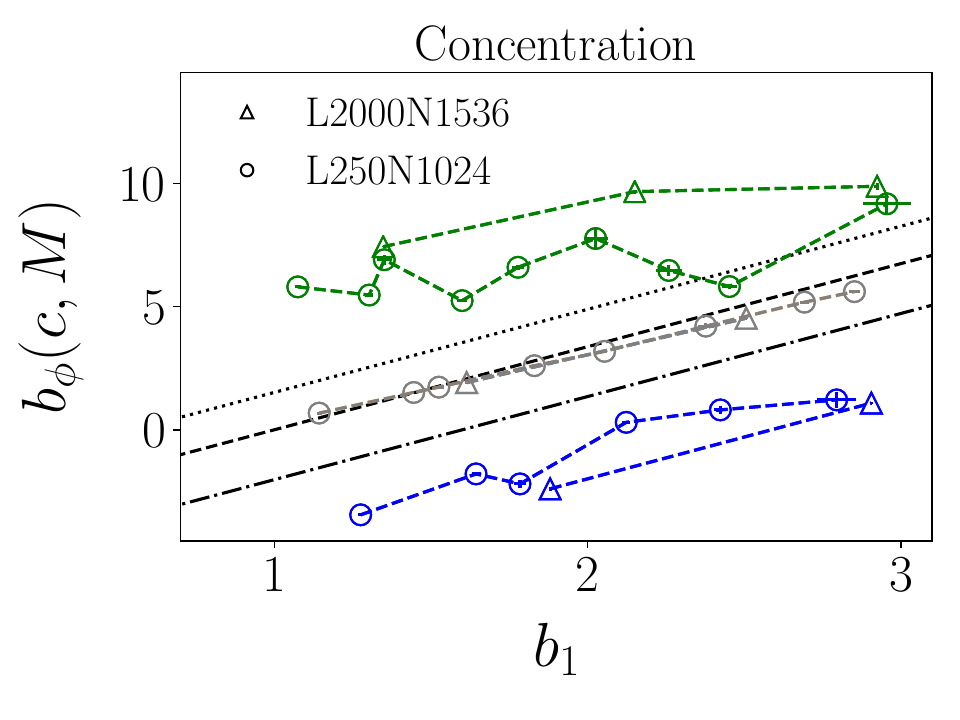}
\includegraphics[scale=0.49]{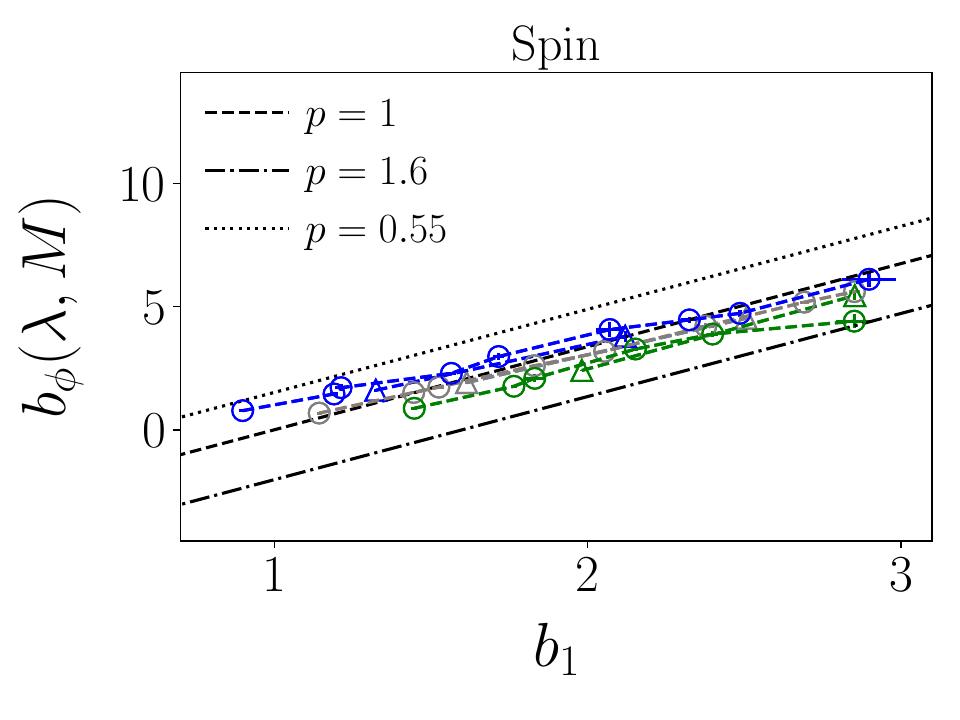}
\includegraphics[scale=0.49]{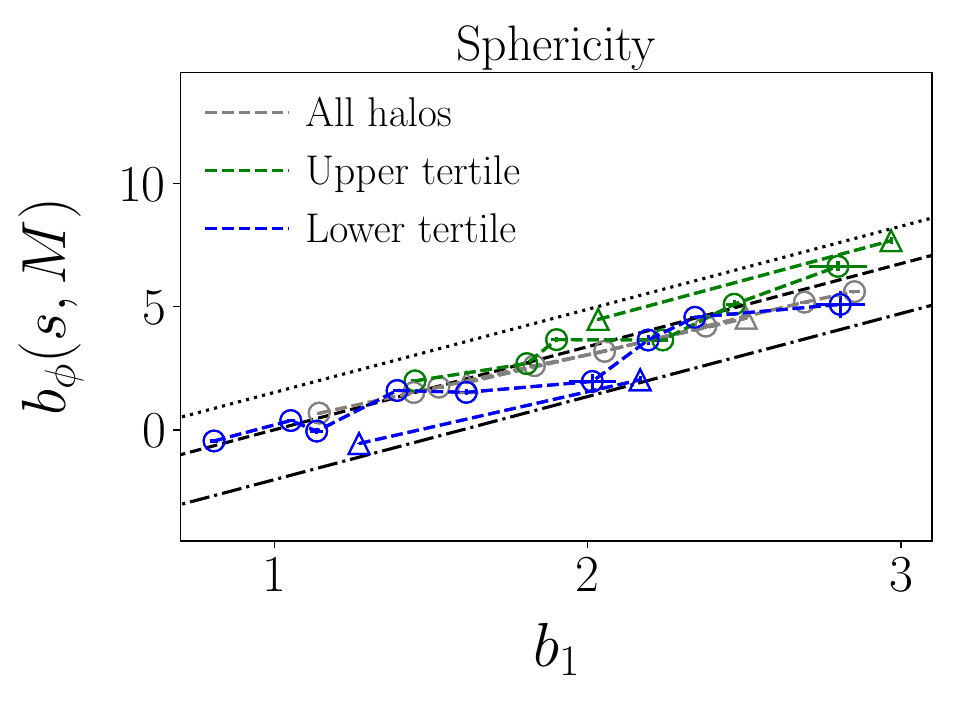}
\caption{Same as Fig.~\ref{fig:bphiofb1ofp_1}, except that the sets of symbols connected with lines show now the $\bphi(b_1)$ relation for halo populations in the same concentration, spin and sphericity tertiles, across the different mass and redshift values (for clarity, the mass and redshift are not marked explicitly, but can be read off from Fig.~\ref{fig:bphiofb1ofp_1}). The results in green/blue show the relations for the halos in the higher/lower secondary property tertiles, and the results in grey show the relation for the whole halo population. The result is shown for the $\seta$ and $\setb$ simulation sets, as labeled. For comparison, the black lines show the relation $\bphi(b_1) = 2\delta_c(b_1-p)$ for $p=1, 1.6, 0.55$, which are relations that have been used in the literature to constrain $\fnl$ (the $p=1$ case is the universality relation). The figure restricts to the range $b_1 \leq 3$ to focus only on the most observationally relevant values.}
\label{fig:bphiofb1ofp_2}
\end{figure}

Concretely, at fixed halo mass, increasing the halo concentration increases $\bphi$ and decreases $b_1$, i.e., it drives the objects upwards and leftwards in the $\bphi-b_1$ plane. The net effect is the development of {\it vertical variations} with concentration that are orthogonal to the direction of the $\bphi(b_1)$ relation of the whole halo population, which is roughly the same as that indicated by the universality relation shown by the dashed black line $\bphi(b_1) = 2\delta_c(b_1-1)$ (cf.~Fig.~\ref{fig:bphiofb1ofp_2} next). On the other hand, although the actual values of $b_1$ and $\bphi$ can be visibly impacted by halo spin and sphericity at fixed halo mass, the impact is similar for both parameters, which roughly preserves their relation (one can discern some impact, especially at higher halo masses, but which is appreciably smaller than the corresponding impact of halo concentration).

Figure \ref{fig:bphiofb1ofp_2} shows a different representation of the halo assembly bias results of Fig.~\ref{fig:bphiofb1ofp_1}. Rather than the lines connecting the bias values in different tertiles in the same mass bin and redshift, now the lines simply connect the results for all halos in a given secondary property tertile, for all mass and redshift values available. The results in green and blue are for the higher and lower tertiles, respectively, and the result in grey shows the relation for all halos in the different mass bins. This figure illustrates again the strong impact that halo concentration has on the $\bphi(b_1)$ relation: the relation for the $33\%$ most (least) concentrated halos lies significantly above (below) that of the whole halo population. Likewise, and as discussed above, spin and sphericity are shown to have some impact on the $\bphi(b_1)$ relation, but the magnitude of the effect is much lower than for concentration. 

In comparison to the universality relation $\bphi(b_1) = 2\delta_c(b_1-1)$ (dashed black line), the upper left panels in Figs.~\ref{fig:bphiofb1ofp_1} and \ref{fig:bphiofb1ofp_2} make it apparent that this simple relation is not an adequate description for halo populations with different concentrations: halos with higher (lower) concentration have a $\bphi(b_1)$ relation that has higher (lower) amplitude than the universality relation. It is also interesting to compare our results against the black dot-dashed line in Fig.~\ref{fig:bphiofb1ofp_2} that shows the variant of the universality relation $\bphi(b_1) = 2\delta_c(b_1-p)$ with $p=1.6$, which was derived by Ref.~\cite{slosar/etal:2008} for halos that have recently merged (see also Ref.~\cite{Reid:2010} for a derivation of another formula that explicitly takes a range of formation redshift $z_f$ into account). The fact that this relation has a lower amplitude than the universality relation is qualitatively in line with our results that lower concentration (younger) halos have lower $\bphi(b_1)$ relations. At face value, the $\bphi(b_1) = 2\delta_c(b_1-1.6)$ variant seems to describe halo populations that have concentrations in between the mean (grey) and the lowest concentration tertile (blue). This relation is commonly adopted in $\fnl$ constraints using quasars \cite{slosar/etal:2008, 2021arXiv210613725M, 2019JCAP...09..010C}, but whether these quasar samples live in halos with such concentrations is currently unknown, and so constraints on $\fnl$ obtained assuming this relation should be interpreted with care. Finally, the dotted line in Fig.~\ref{fig:bphiofb1ofp_2} shows the variant $\bphi(b_1) = 2\delta_c(b_1-p)$ with $p=0.55$, that (i) Ref.~\cite{2020JCAP...12..013B} found to roughly describe the relation of galaxies selected by stellar mass in the IllustrisTNG galaxy formation model, and (ii) which seems to describe halos with concentrations in between the mean and the higher tertile. In future work, it would be interesting to link these two results together by inspecting the typical concentrations of the host halos of IllustrisTNG galaxies.\footnote{
We note in passing that in addition to the halo occupation distribution (HOD) of the galaxies, there is another physical effect that is needed to link halo to galaxy bias, which is the response of the HOD to the large-scale perturbations. Concretely, we can write the $\bphi^{\rm g}$ parameter of the galaxies as (cf.~Ref.~\cite{Voivodic:2020bec} for the derivation, and note that an analogous expression can be written for any bias parameter)
\bq\label{eq:hodresponse}
\bphi^g = \frac{1}{n_{g}}\int {\rm d}M \int {\rm d}c\ n_{h}(M, c) N_g(M, c) \left(\bphi^h(M, c) + R_\phi^{N_g}(M, c)\right),
\eq
where $n_{g}$ is the galaxy number density, $n_{h}(M, c)$ is the mean number density of halos, $N_g(M,c)$ is the mean number of galaxies that live in halos of mass $M$ and concentration $c$, and $\bphi^h$ is the bias parameter of the halos. The second term in the brackets is defined as $R_\phi^{N_g} = \partial \ln N_g(M,c)/\partial{(\fnl\phi)}$, and describes the response of the HOD to long-wavelength $\fnl\phi$ perturbations. That is, even if one knows exactly how galaxies occupy halos on average, i.e.~we know the function $N_g(M, c)$, one still needs to know how this occupation is modulated by $\fnl\phi$ perturbations (or equivalently changes to $\mathcal{A}_s$) to fully determine the bias parameter of the galaxies $\bphi^g$. This physical effect is often ignored in the literature, but Ref.~\cite{Voivodic:2020bec} showed it can be important.
}

Before proceeding, we note that our halo assembly bias results in this paper for the $\bphi(b_1)$ relation find interesting connection points with the halo assembly bias results of Ref.~\cite{Lazeyras:2021dar} for the $b_2(b_1)$ and $b_{K^2}(b_1)$ relations, where $b_2$ is the second-order density bias parameter and $b_{K^2}$ is the leading-order halo tidal bias parameter. Concretely, Ref.~\cite{Lazeyras:2021dar} studied the assembly bias signal in these two bias parameters and their relations with $b_1$ for the same three secondary halo properties that we study in this paper. The authors found that the $b_2(b_1)$ relation remains remarkably weakly affected by the assembly bias signal, but the $b_{K^2}(b_1)$ can be significantly affected by assembly bias for all three secondary halo properties. This can be contrasted with our results here for the $\bphi(b_1)$ relation, which is strongly affected by concentration, but weakly affected by spin and sphericity. The bias parameters $b_1$, $b_2$, $b_{K^2}$ and $\bphi$ describe different aspects of halo formation, which makes the differential impact of halo assembly bias on them interesting to study from a cosmic structure formation physics point of view. The way these assembly bias effects on the halo bias relations propagate to the corresponding relations for visible tracers like galaxies \cite{Voivodic:2020bec, Barreira:2021ukk} and neutral hydrogen \cite{2022JCAP...04..057B, 2022arXiv220712398O} would be also interesting to investigate in future work.


\section{Consequences for $f_{\rm NL}$ constraints}
\label{sec:fnl}

\begin{figure}[t]
\centering
\includegraphics[width=\textwidth]{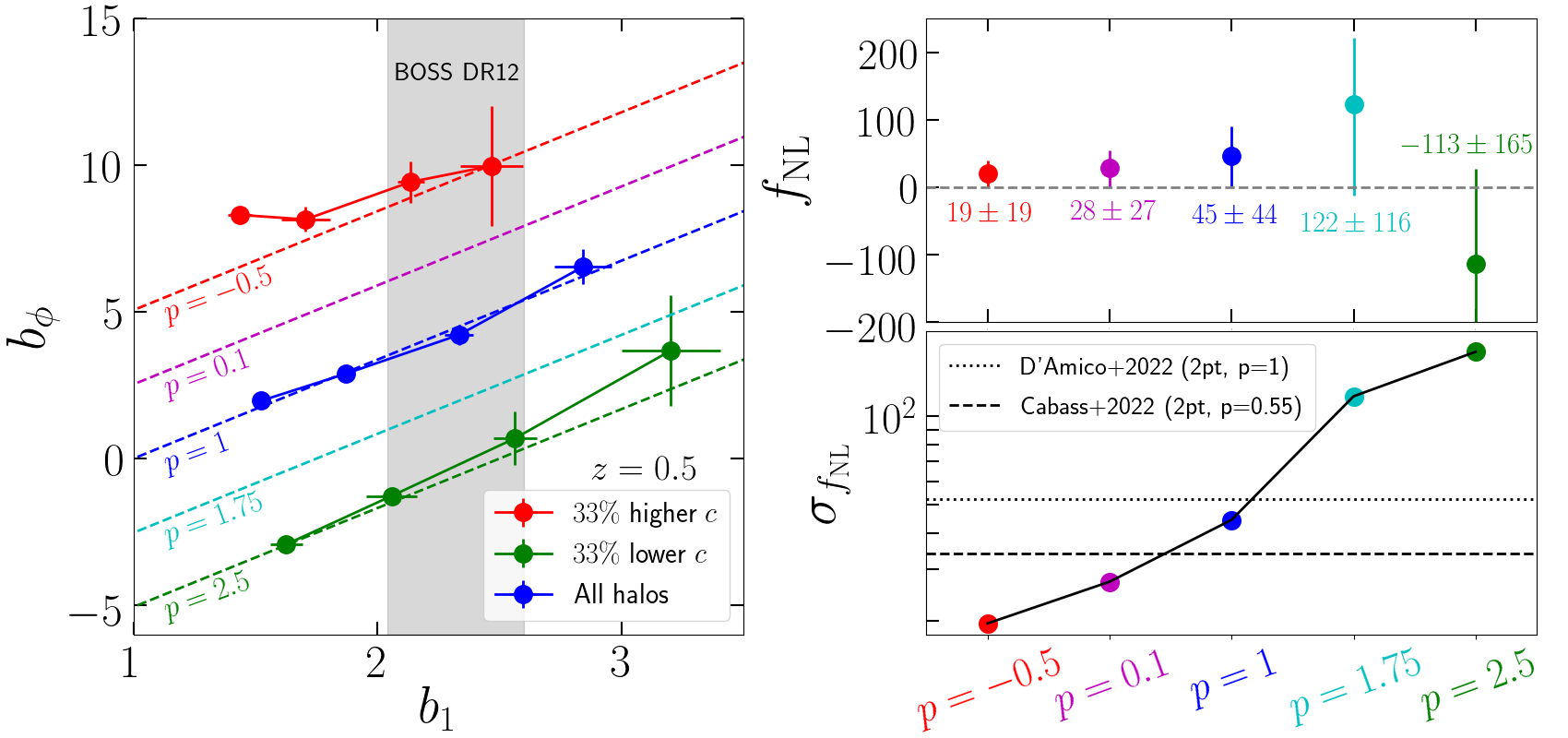}
\caption{The potential consequences of halo assembly bias for observational constraints on $\fnl$ using galaxy data. In the left panel, the blue symbols show the $\bphi(b_1)$ relation for halos in total mass bins. The green and red symbols show the same, but for the halos in the lowest and highest concentration tertile in the same mass bins. For each of these cases, from left to right, the symbols correspond to halos with $M = \left\{1,2,4,8\right\}\times10^{13} M_{\odot}/h$ at $z = 0.5$. This corresponds roughly to the mass and mean redshift covered by galaxies in the BOSS DR12 survey. The result is from the $\setc$ simulation set. The vertical grey band marks the constraints on $b_1$ found in the power spectrum and bispectrum analysis of Ref.~\cite{2022PhRvD.105d3517P} for BOSS DR12 galaxies (see their Table VII). The right panels shows the constraints on $\fnl$ using the BOSS DR12 galaxy power spectrum for different values of $p$ in the parametrization $\bphi(b_1) = 2\delta_c(b_1-p)$. The corresponding $\bphi(b_1)$ relations for the five values of $p$ considered is shown by the dashed lines on the left. For reference, in the lower right panel, the horizontal lines show the constraints obtained by Refs.~\cite{2022arXiv220111518D, 2022arXiv220401781C} for the same galaxies in the power-spectrum-only part of their analysis.}
\label{fig:boss}
\end{figure}

In this section, we wish to roughly estimate the impact that halo assembly bias can have on $\fnl$ constraints, taking as a working-case the galaxy power spectrum of BOSS DR12 galaxies. The symbols in the left panel of Fig.~\ref{fig:boss} show the $\bphi(b_1)$ relation for all halos in blue, and halos in the lowest and highest concentration tertiles in green and red, respectively. The four symbols cover halo masses in the range $M \in \left[1; 8\right]\times10^{13} M_{\odot}/h$ at $z=0.5$, which is broadly representative for BOSS DR12 galaxies \cite{2011ApJ...728..126W, 2013MNRAS.429...98P, 2016MNRAS.455.1553R}; the result is obtained using the $\setc$ simulation set. The dashed lines show 5 different $\bphi(b_1)$ relations parametrized by different values of $p \in \left[-0.5, 2.5\right]$ in the equation $\bphi(b_1) = 2\delta_c(b_1 - p)$, chosen to cover the possibilities bracketed by the simulation relations shown.

The way BOSS galaxies populate halos as a function of concentration is currently not well known, and so it is hard to pinpoint which relation is closest to that of these galaxies' host halos. Using stacked lensing profiles for the subset of BOSS galaxies that overlap with the CFHT Stripe 82 Survey, Ref.~\cite{2017ApJ...840..104S} constrained the typical host halo concentration to be in the range $c \sim 2.5 - 5.8$.\footnote{This corresponds to the range of $c_{200c}$ values allowed by the $1\sigma$ bounds in their Tab.~1.} In our $\setc$ simulations at $z=0.5$, the mean concentration of the halos in the lowest and highest tertiles is $\approx 4$ and $\approx 5.5$, respectively, for halo masses $M = 2-4\times 10^{13} M_{\odot}/h$. Noting that a detailed quantitative comparison cannot be performed as our snapshots do not cover exactly the redshift range of the BOSS galaxies and the concentration-mass relation depends also on cosmology, we will use the fact that the range of concentration values spanned by our tertiles is smaller than the current uncertainty on the concentration of the host halos of BOSS galaxies to justify that it is currently plausible to roughly  expect the $\bphi(b_1)$ relation of the BOSS galaxies to be bracketed by the values of $p$ marked in Fig.~\ref{fig:boss}.

The constraints on $\fnl$ are shown in the right panels of Fig.~\ref{fig:boss}; the actual $\fnl$ bounds are in the upper panel, and the lower panel isolates the impact on the error bar $\sigma_{\fnl}$. These constraints were obtained using the same methodology of Ref.~\cite{2022arXiv220505673B}, assuming the following linear redshift-space power spectrum model
\bq\label{eq:modelPkmu}
P_{gg}(k,\mu) = \Bigg[\left(b_1 + f\mu^2\right)^2 + \frac{2\left(b_1 + f\mu^2\right)\bphi\fnl}{\mathcal{M}(k)} + \frac{(\bphi\fnl)^2}{\mathcal{M}(k)^2}\Bigg] P_{mm}(k) + \frac{\alpha_P}{\bar{n}_g},
\eq
where $\mu$ is the cosine of the angle between the line-of-sight direction and the wavevector $\vk$, $\bar{n}_g$ is the galaxy number density, $\alpha_P$ is a parameter that quantifies departures of the shot-noise from the Poisson expectation and $f = {\rm dln}D/{\rm dln}a$ is the linear growth factor. We use this model to compute the monopole ($\ell = 0$) and quadrupole ($\ell = 2$) moments defined as 
\bq\label{eq:Pkmulti}
P_{gg}^{\ell}(k) = \frac{2\ell+1}{2} \int_{-1}^1 {\rm d}\mu P_{gg}(k,\mu) L_{\ell}(\mu), 
\eq
where $L_{\ell}(\mu)$ are Legendre polynomials, which we fit to the measurements obtained with the method of Ref.~\cite{2021PhRvD.103j3504P} for four galaxy samples in BOSS DR12 ($z_1 = 0.38$ and $z_3 = 0.61$ in the north and south galactic caps).\footnote{These are publicly available at \url{https://github.com/oliverphilcox/Spectra-Without-Windows}. Note that these galaxy samples overlap with the samples known as LOWZ and CMASS, but are not exactly the same.} Except for $\fnl$, we keep all cosmological parameters fixed\footnote{Specifically, we consider a standard flat $\Lambda$CDM model with parameters:  $h = 0.6778$, $\Omega_b h^2 = 0.02268$, $\Omega_c h^2 = 0.1218$, $n_s = 0.9649$ and $\sigma_8 = 0.75$. This is the best-fitting cosmology in the left-hand side of Table VII of Ref.~\cite{2022PhRvD.105d3517P}.}, and for each galaxy sample, we fit also for the values of $b_1$ and Poisson shot-noise amplitude; we assume Gaussian priors for $b_1$ using the 1-loop power spectrum and  bispectrum constraints in Table VII of Ref.~\cite{2022PhRvD.105d3517P}, which are for the same cosmology. We adopt a conservative choice for the maximum wavenumber $k_{\rm max} = 0.05 h/{\rm Mpc}$ to ensure the validity of linear theory. Note that while the constraints on $\fnl$ are dominated by the largest scales, extending the analysis to smaller scales is useful in general to constrain the cosmological parameters and $b_1$; though here we keep the former fixed and assume priors for $b_1$. We stress that our goal is not to derive the best possible constraints, but just to illustrate the impact of the different $\bphi(b_1)$ relations. The covariance matrix was obtained using the power spectra from the ensemble of 2048 MultiDark-Patchy galaxy mock catalogues \cite{2016MNRAS.456.4156K, 2016MNRAS.460.1173R}. We refer the reader to Ref.~\cite{2022arXiv220505673B} for more details about the constraint methodology and its validation.

The $p = -0.5$ and $p = 0.1$ cases yield the largest values of $\bphi$ ($\bphi \approx 10$ and $\bphi \approx 7$ for the typical BOSS DR12 $b_1$ values, respectively), and consequently, give the tightest constraints on $\fnl$ shown in Fig.~\ref{fig:boss}: $\fnl = 19 \pm 19\ (1\sigma)$ and $\fnl = 28 \pm 27\ (1\sigma)$, respectively. For comparison, both these constraints are tighter than the power-spectrum-only results of Refs.~\cite{2022arXiv220111518D, 2022arXiv220401781C} marked by the horizontal lines in the lower right panel of Fig.~\ref{fig:boss}.\footnote{Reference~\cite{2022arXiv220111518D} assumed $p=1$, which is the universality relation. Reference~\cite{2022arXiv220401781C} assumed $p=0.55$ inspired by the results of Ref.~\cite{2020JCAP...12..013B}, who found this for stellar mass selected galaxies in simulations of the IllustrisTNG model.} That is, if BOSS DR12 galaxies happen to populate halos that are more concentrated than average, then existing $\fnl$ constrains using BOSS DR12 galaxies may currently underestimate the true constraining power of the data. On the other hand, should these galaxies preferentially reside in lower concentration halos, this brings their $\bphi$ values closer to zero and weakens the constraints on $\fnl$. Concretely, the $p = 2.5$ case assumes that BOSS galaxies inhabit a population of halos in the lowest concentration tertile at $z=0.5$. This gives a constraint on $\fnl$ ($\sigma_{\fnl}=165$) that is weaker by a factor of $3.8$ compared to assuming $p=1$ ($\sigma_{\fnl}=44$), which is close to the global halo population. Note that for the typical $b_1$ values of BOSS DR12 galaxies, the two concentration tertiles bracket situations in which $\bphi$ can be a small number and yield very weak constraints on $\fnl$. This shows how relatively small changes in $\bphi$ can have a dramatic effect on $\fnl$ constraints.

Considerations about how galaxies populate halos as a function of mass and concentration are currently ignored when deciding which $\bphi(b_1)$ relation to assume in $\fnl$ constraints, but our simple analysis here highlights that this needs to be corrected in the future given the strong impact this can have on the $\fnl$ bounds. In particular, given the current uncertainties on the concentration-mass relation of the host halos of BOSS DR12 galaxies, Fig.~\ref{fig:boss} shows that it is premature to claim a certain constraining power on $\fnl$ from these data. Note that this message is not specific to the BOSS DR12 samples, and applies generically to any tracer of the large-scale structure that can be used to constrain $\fnl$ (including emission line galaxies, quasars, neutral Hydrogen, etc.). The discussion in this section adds on to other analyses and discussions in the literature \cite{Reid:2010, 2019JCAP...09..010C, 2020JCAP...12..031B, 2020JCAP...12..013B, 2022JCAP...01..033B, 2022arXiv220505673B} that have been calling attention to the need to improve our current knowledge about the $\bphi(b_1)$ relation in order to be able to reliably and competitively constrain $\fnl$ using galaxy data.

Along the same lines, it is important to stress that, in fact, the significance of detection of $\fnl \neq 0$ is not affected by different assumptions on $\bphi$. Indeed, the various constraints shown on the right of Fig.~\ref{fig:boss} all {\it detect} $\fnl \neq 0$ at $\approx 1\sigma$. In assessing the significance of detection, however, the appropriate thing to do is to constrain the parameter combination $\fnl\bphi$, and not $\fnl$ (see Refs.~\cite{2020JCAP...12..031B, 2020JCAP...12..013B} for discussions about the pros and cons of this approach). Constraints on $\fnl\bphi$ do not strictly require $\bphi$ priors, but some knowledge of the $\bphi(b_1)$ relation is still useful to target samples with $|\bphi| \gg 1$ to maximize the signal-to-noise of the detection. In fact, a general characterization of trends on the $\bphi-b_1$ plane (as opposed to very precise knowledge of the $\bphi(b_1)$ relation) can prove sufficient to identify which tracers are best in terms of significance of detection. However, should future constraints on $\fnl\bphi$ remain compatible with zero like ours here, then the precision and accuracy of the $\bphi$ priors become critical to determine the upper bounds on $|\fnl|$.


\section{Summary \& Conclusion}
\label{sec:concl} 

Observational constraints on the local PNG parameter $\fnl$ obtained using galaxy data come primarily from signatures associated with the scale-dependent bias effect. The leading-order signature appears in the large-scale galaxy power spectrum and it is $\propto b_1\bphi\fnl/k^2$, where $b_1$ and $\bphi$ are the galaxy bias parameters associated with mass density and primordial gravitational potential perturbations, respectively. While the value of $b_1$ can be directly estimated from the data, the same is not true for $\bphi$, unless prior knowledge on $\fnl$ is used. Thus, in order to constrain $\fnl$, one needs to make assumptions about $\bphi$. The standard approach in the literature is to fix $\bphi$ in terms of $b_1$ by assuming a tight relation between the two. This relation is often taken to be the universality relation $\bphi(b_1) = 2\delta_c(b_1 - 1)$ (or certain simplified variants thereof), but there is no compelling reason to expect this to describe the $\bphi(b_1)$ relation of tracers in real-life observational surveys. Indeed, a number of recent works based on hydrodynamical simulations of galaxy formation have revealed that the $\bphi(b_1)$ relation can vary sensitively between different types of tracers of the large-scale structure \cite{2020JCAP...12..013B, 2022JCAP...01..033B, 2022JCAP...04..057B}. Without the ability to place good priors on the $\bphi(b_1)$ relation it may not be possible to use galaxy data to competitively constrain $\fnl$ \cite{2022arXiv220505673B}, which motivates works like the present one focused on studying it.

Concretely, in this paper we studied halo assembly bias in $b_1$ and $\bphi$ and its impact on the $\bphi(b_1)$ relation. This is an important step towards a robust understanding of the same relation for galaxies, because different galaxy types may preferentially reside in halos with distinct properties (at fixed halo mass) and thus inherit the assembly bias signal of their host halos. We used a series of gravity-only simulations to investigate how the $\bphi(b_1)$ relation depends on halo concentration, spin and sphericity, at fixed halo mass. The bias parameter $b_1$ was estimated using the large-scale limit of the ratio of the halo-matter to matter-matter power spectrum, and the bias parameter $\bphi$ was estimated using separate universe simulations with varying normalization amplitude $\sigma_8$ (equivalently the PBS argument, cf.~Sec.~\ref{sec:biasest}). The halo assembly bias signal in $\bphi$ had been investigated previously in detail by Ref.~\cite{Reid:2010}, who showed that halos with the same mass but different formation times can have significantly different values of $\bphi$. Our work here can be regarded as an extension of that by Ref.~\cite{Reid:2010} to study the assembly bias effect on the $b_\phi(b_1)$ \emph{relation} and to consider also halo concentration, spin and sphericity as additional secondary halo properties. Our main results can be summarized as follows:

\begin{itemize}

\item We detect the halo assembly bias signal in $\bphi$ for all three secondary halo properties considered. In particular, at fixed halo mass, our results show that $\bphi$ typically grows with halo concentration, spin and sphericity (cf.~Fig.~\ref{fig:bofp}). Noting that higher concentration halos are typically halos that formed earlier, our concentration results therefore qualitatively agree with the previous finding of Ref.~\cite{Reid:2010} that older halos have larger $\bphi$.

\item The halo assembly bias signal from halo spin and sphericity is similar for $b_1$ and $\bphi$, and roughly preserves the $\bphi(b_1)$ relation of the full halo sample (cf.~Figs.~\ref{fig:bphiofb1ofp_1} and \ref{fig:bphiofb1ofp_2}). By contrast, varying the halo concentration shifts $b_1$ and $\bphi$ in opposite directions and impacts $\bphi$ appreciably more than $b_1$. The net result is a strong modification of the $\bphi(b_1)$ relation, which acquires a higher (lower) amplitude for higher (lower) concentration halos (cf.~Fig.~\ref{fig:bphiofb1ofp_2}). This differential impact of halo concentration, spin and sphericity on the $b_1$ and $\bphi$ parameters is interesting also from the point of view of the physics of cosmic structure formation, which would be interesting to investigate in future work.

\end{itemize}

The strong impact of halo concentration on the $\bphi(b_1)$ relation has important ramifications for $\fnl$ constraints, which currently do not take this effect into account. In Sec.~\ref{sec:fnl}, we presented a simple estimate of the importance of this to $\fnl$ constraints using BOSS DR12 galaxies, which we found can be significant. Assuming that BOSS galaxies are a representative sample of the whole halo concentration distribution, we obtain $\sigma_{\fnl} = 44$. Assuming instead that BOSS galaxies occupy the $33\%$ most and least concentrated halos, we obtained $\sigma_{\fnl} = 19$ and $\sigma_{\fnl} = 165$, respectively (cf.~Fig.~\ref{fig:boss}). The range of values of $\bphi$ spanned by these two concentration tertiles can cross zero, and so under this simplified analysis it is also possible to identify a population of halos with $|\bphi \ll 1|$ that would yield extremely weak constraints on $\fnl$. Further, observational constraints using quasars do often quote results for a variant of the universality relation $\bphi(b_1) = 2\delta_c(b_1 - 1.6)$ that attempts to describe objects that have recently merged. Figure \ref{fig:bphiofb1ofp_2} shows that this relation describes the $\bphi(b_1)$ relation of halos with concentrations between the mean and the lower tertile, but whether quasars occupy these or other halo populations is currently unknown. The takeaway is that our uncertain knowledge of the concentration of the host halos of observed tracers currently prevents us from determining the constraining power of galaxy data on $\fnl$.

Although the strong sensitivity of the $\bphi(b_1)$ relation to the details of the halo and galaxy populations may seem alarming and challenging, this in principle brings with it the opportunity to optimize galaxy selection strategies in surveys to simply detect $\fnl \neq 0$. Constraints on $\fnl\bphi$ do not require $\bphi$ priors, and detecting $\fnl\bphi \neq 0$ would still allow to rule out single field inflation. In contrast to $\fnl$ constraints that require accurate and precise priors on $\bphi$, in $\fnl\bphi$ constraints even a rough understanding of the $\bphi(b_1)$ relation of the galaxies in surveys could go a long way in identifying what types of tracers are best to detect $\fnl \neq 0$. For example, the results of Ref.~\cite{Maccio:2006wpz} suggest that the surface brightness of disk galaxies may offer a way to select host halos with certain concentration values. Concretely, the correlation between halo spin and concentration, along with centrifugal considerations, imply that, at fixed halo mass, high (low) surface brightness disks reside in more (less) concentrated halos, and may as a result have a higher (lower) amplitude of the $\bphi(b_1)$ relation. Alternatively, although existing lensing-based estimates of the concentration of individual galaxies' host halos are currently very noisy, it is interesting to investigate whether in the future these can be made precise enough for selection cuts for $\fnl\bphi$ constraints; note that for a simple split into high/low concentration objects, the precision of the individual concentration estimates may not be as important. Furthermore, Ref.~\cite{2022JCAP...01..033B} found using hydrodynamical simulations that galaxies with lower black hole mass accretion rate (and by proxy, lower quasar luminosity) tend to have larger values of $\bphi$ at fixed $b_1$ (cf.~their Fig.~3), and as a result are more sensitive to $\fnl$. The investigation of which galaxy types are best (or worst) to detect $\fnl\bphi$ through their $|\bphi|$ values is an important line of research that is interesting to pursue further in future work. 

\acknowledgments{TL acknowledges support from the European Research Council (ERC) under the European Union's Horizon 2020 research and innovation program grant agreement No 864361. AB acknowledges support from the  Excellence  Cluster  ORIGINS  which  is  funded  by  the  Deutsche  Forschungsgemeinschaft  (DFG, German Research Foundation) under Germany’s Excellence Strategy - EXC-2094-390783311.  FS acknowledges support from the Starting Grant (ERC-2015-STG 678652) “GrInflaGal” of the European Research Council.  VD acknowledges support by the Israel Science Foundation (ISF) grants no 2562/20, and would also like to thank the Max-Planck Institute for Astrophysics (Garching) for hospitality during the completion of this work. The simulations used in this work and their numerical analysis was  done  on  the  Freya and Cobra supercomputers of the Max Planck Computing and Data Facility (MPCDF). This research was supported by the Munich Institute for Astro, -Particle and BioPhysics (MIAPbP) which is funded by the Deutsche Forschungsgemeinschaft (DFG, German Research Foundation) under Germany's Excellence Strategy – EXC-2094 – 390783311. Part of this work was performed at the Aspen Center for Physics, which is supported by National Science Foundation grant PHY-1607611.}


\FloatBarrier
\bibliography{references}

\providecommand{\href}[2]{#2}\begingroup\raggedright\begin{thebibliography}{10}

\bibitem{2001PhRvD..63f3002K}
E.~{Komatsu} and D.~N. {Spergel}, {\it {Acoustic signatures in the primary
  microwave background bispectrum}},  {\em \prd} {\bf 63} (Mar., 2001)
  063002--+, [\href{http://arxiv.org/abs/astro-ph/0}{{\tt astro-ph/0}}].

\bibitem{maldacena:2003}
J.~{Maldacena}, {\it {Non-gaussian features of primordial fluctuations in
  single field inflationary models}},  {\em Journal of High Energy Physics}
  {\bf 5} (May, 2003) 13, [\href{http://arxiv.org/abs/astro-ph/0}{{\tt
  astro-ph/0}}].

\bibitem{2004JCAP...10..006C}
P.~{Creminelli} and M.~{Zaldarriaga}, {\it {A single-field consistency relation
  for the three-point function}},  {\em \jcap} {\bf 2004} (Oct., 2004) 006,
  [\href{http://arxiv.org/abs/astro-ph/0407059}{{\tt astro-ph/0407059}}].

\bibitem{2011JCAP...11..038C}
P.~{Creminelli}, G.~{D'Amico}, M.~{Musso}, and J.~{Nore{\~n}a}, {\it {The (not
  so) squeezed limit of the primordial 3-point function}},  {\em \jcap} {\bf
  2011} (Nov., 2011) 038, [\href{http://arxiv.org/abs/1106.1462}{{\tt
  arXiv:1106.1462}}].

\bibitem{Tanaka:2011aj}
T.~Tanaka and Y.~Urakawa, {\it {Dominance of gauge artifact in the consistency
  relation for the primordial bispectrum}},  {\em JCAP} {\bf 1105} (2011) 014,
  [\href{http://arxiv.org/abs/1103.1251}{{\tt arXiv:1103.1251}}].

\bibitem{2020A&A...641A...9P}
{Planck Collaboration}, {\it {Planck 2018 results. IX. Constraints on
  primordial non-Gaussianity}},  {\em \aap} {\bf 641} (Sept., 2020) A9,
  [\href{http://arxiv.org/abs/1905.05697}{{\tt arXiv:1905.05697}}].

\bibitem{2014arXiv1412.4872D}
O.~{Dor{\'e}}, J.~{Bock}, M.~{Ashby}, P.~{Capak}, A.~{Cooray}, R.~{de Putter},
  T.~{Eifler}, N.~{Flagey}, Y.~{Gong}, and S.~{Habib}, {\it {Cosmology with the
  SPHEREX All-Sky Spectral Survey}},  {\em arXiv e-prints} (Dec, 2014)
  arXiv:1412.4872, [\href{http://arxiv.org/abs/1412.4872}{{\tt
  arXiv:1412.4872}}].

\bibitem{2014arXiv1412.4671A}
M.~{Alvarez} and {et al}, {\it {Testing Inflation with Large Scale Structure:
  Connecting Hopes with Reality}},  {\em arXiv e-prints} (Dec, 2014)
  arXiv:1412.4671, [\href{http://arxiv.org/abs/1412.4671}{{\tt
  arXiv:1412.4671}}].

\bibitem{2017PhRvD..95l3513D}
R.~{de Putter} and O.~{Dor{\'e}}, {\it {Designing an inflation galaxy survey:
  How to measure {\ensuremath{\sigma}}($f_{NL}) \sim 1$ using scale-dependent
  galaxy bias}},  {\em \prd} {\bf 95} (Jun, 2017) 123513,
  [\href{http://arxiv.org/abs/1412.3854}{{\tt arXiv:1412.3854}}].

\bibitem{2019Galax...7...71B}
M.~{Biagetti}, {\it {The Hunt for Primordial Interactions in the Large-Scale
  Structures of the Universe}},  {\em Galaxies} {\bf 7} (Aug., 2019) 71,
  [\href{http://arxiv.org/abs/1906.12244}{{\tt arXiv:1906.12244}}].

\bibitem{2021arXiv210609713S}
N.~{Sailer}, E.~{Castorina}, S.~{Ferraro}, and M.~{White}, {\it {Cosmology at
  high redshift -- a probe of fundamental physics}},  {\em arXiv e-prints}
  (June, 2021) arXiv:2106.09713, [\href{http://arxiv.org/abs/2106.09713}{{\tt
  arXiv:2106.09713}}].

\bibitem{2022arXiv220307506F}
S.~{Ferraro}, N.~{Sailer}, A.~{Slosar}, and M.~{White}, {\it {Snowmass2021
  Cosmic Frontier White Paper: Cosmology and Fundamental Physics from the
  three-dimensional Large Scale Structure}},  {\em arXiv e-prints} (Mar., 2022)
  arXiv:2203.07506, [\href{http://arxiv.org/abs/2203.07506}{{\tt
  arXiv:2203.07506}}].

\bibitem{2022arXiv220308128A}
A.~{Ach{\'u}carro}, M.~{Biagetti}, M.~{Braglia}, G.~{Cabass}, E.~{Castorina},
  R.~{Caldwell}, X.~{Chen}, W.~{Coulton}, R.~{Flauger}, J.~{Fumagalli}, M.~M.
  {Ivanov}, H.~{Lee}, A.~{Maleknejad}, P.~D. {Meerburg}, A.~{Moradinezhad
  Dizgah}, G.~A. {Palma}, S.~{Renaux-Petel}, G.~L. {Pimentel}, B.~{Wallisch},
  B.~D. {Wandelt}, L.~T. {Witkowski}, and W.~L. {Kimmy Wu}, {\it {Inflation:
  Theory and Observations}},  {\em arXiv e-prints} (Mar., 2022)
  arXiv:2203.08128, [\href{http://arxiv.org/abs/2203.08128}{{\tt
  arXiv:2203.08128}}].

\bibitem{slosar/etal:2008}
A.~{Slosar}, C.~{Hirata}, U.~{Seljak}, S.~{Ho}, and N.~{Padmanabhan}, {\it
  {Constraints on local primordial non-Gaussianity from large scale
  structure}},  {\em \jcap} {\bf 8} (Aug., 2008) 31--+,
  [\href{http://arxiv.org/abs/0805.3580}{{\tt arXiv:0805.3580}}].

\bibitem{mcdonald:2008}
P.~{McDonald}, {\it {Primordial non-Gaussianity: Large-scale structure
  signature in the perturbative bias model}},  {\em \prd} {\bf 78} (Dec., 2008)
  123519--+, [\href{http://arxiv.org/abs/0806.1061}{{\tt arXiv:0806.1061}}].

\bibitem{giannantonio/porciani:2010}
T.~{Giannantonio} and C.~{Porciani}, {\it {Structure formation from
  non-Gaussian initial conditions: Multivariate biasing, statistics, and
  comparison with N-body simulations}},  {\em \prd} {\bf 81} (Mar., 2010)
  063530--+, [\href{http://arxiv.org/abs/0911.0017}{{\tt arXiv:0911.0017}}].

\bibitem{2011JCAP...04..006B}
T.~{Baldauf}, U.~{Seljak}, and L.~{Senatore}, {\it {Primordial non-Gaussianity
  in the bispectrum of the halo density field}},  {\em Journal of Cosmology and
  Astro-Particle Physics} {\bf 2011} (Apr., 2011) 006,
  [\href{http://arxiv.org/abs/1011.1513}{{\tt arXiv:1011.1513}}].

\bibitem{assassi/baumann/schmidt}
V.~{Assassi}, D.~{Baumann}, and F.~{Schmidt}, {\it {Galaxy bias and primordial
  non-Gaussianity}},  {\em \jcap} {\bf 12} (Dec., 2015) 043,
  [\href{http://arxiv.org/abs/1510.03723}{{\tt arXiv:1510.03723}}].

\bibitem{Desjacques:2016}
V.~Desjacques, D.~Jeong, and F.~Schmidt, {\it {Large-Scale Galaxy Bias}},  {\em
  Phys. Rept.} {\bf 733} (2018) 1--193,
  [\href{http://arxiv.org/abs/1611.09787}{{\tt arXiv:1611.09787}}].

\bibitem{dalal/etal:2008}
N.~{Dalal}, O.~{Dor{\'e}}, D.~{Huterer}, and A.~{Shirokov}, {\it {Imprints of
  primordial non-Gaussianities on large-scale structure: Scale-dependent bias
  and abundance of virialized objects}},  {\em \prd} {\bf 77} (June, 2008)
  123514--+, [\href{http://arxiv.org/abs/0710.4560}{{\tt arXiv:0710.4560}}].

\bibitem{2021arXiv210613725M}
E.-M. {Mueller} and {et al}, {\it {The clustering of galaxies in the completed
  SDSS-IV extended Baryon Oscillation Spectroscopic Survey: Primordial
  non-Gaussianity in Fourier Space}},  {\em arXiv e-prints} (June, 2021)
  arXiv:2106.13725, [\href{http://arxiv.org/abs/2106.13725}{{\tt
  arXiv:2106.13725}}].

\bibitem{2019JCAP...09..010C}
E.~{Castorina}, N.~{Hand}, U.~{Seljak}, F.~{Beutler}, C.-H. {Chuang},
  C.~{Zhao}, H.~{Gil-Mar{\'\i}n}, W.~J. {Percival}, A.~J. {Ross}, P.~D. {Choi},
  K.~{Dawson}, A.~{de la Macorra}, G.~{Rossi}, R.~{Ruggeri}, D.~{Schneider},
  and G.-B. {Zhao}, {\it {Redshift-weighted constraints on primordial
  non-Gaussianity from the clustering of the eBOSS DR14 quasars in Fourier
  space}},  {\em \jcap} {\bf 2019} (Sep, 2019) 010,
  [\href{http://arxiv.org/abs/1904.08859}{{\tt arXiv:1904.08859}}].

\bibitem{2022arXiv220615450C}
W.~R. {Coulton}, F.~{Villaescusa-Navarro}, D.~{Jamieson}, M.~{Baldi},
  G.~{Jung}, D.~{Karagiannis}, M.~{Liguori}, L.~{Verde}, and B.~D. {Wandelt},
  {\it {Quijote PNG: The information content of the halo power spectrum and
  bispectrum}},  {\em arXiv e-prints} (June, 2022) arXiv:2206.15450,
  [\href{http://arxiv.org/abs/2206.15450}{{\tt arXiv:2206.15450}}].

\bibitem{2022arXiv220111518D}
G.~{D'Amico}, M.~{Lewandowski}, L.~{Senatore}, and P.~{Zhang}, {\it {Limits on
  primordial non-Gaussianities from BOSS galaxy-clustering data}},  {\em arXiv
  e-prints} (Jan., 2022) arXiv:2201.11518,
  [\href{http://arxiv.org/abs/2201.11518}{{\tt arXiv:2201.11518}}].

\bibitem{2022arXiv220401781C}
G.~{Cabass}, M.~M. {Ivanov}, O.~H.~E. {Philcox}, M.~{Simonovi{\'c}}, and
  M.~{Zaldarriaga}, {\it {Constraints on Multi-Field Inflation from the BOSS
  Galaxy Survey}},  {\em arXiv e-prints} (Apr., 2022) arXiv:2204.01781,
  [\href{http://arxiv.org/abs/2204.01781}{{\tt arXiv:2204.01781}}].

\bibitem{grossi/etal:2009}
M.~{Grossi}, L.~{Verde}, C.~{Carbone}, K.~{Dolag}, E.~{Branchini},
  F.~{Iannuzzi}, S.~{Matarrese}, and L.~{Moscardini}, {\it {Large-scale
  non-Gaussian mass function and halo bias: tests on N-body simulations}},
  {\em \mnras} {\bf 398} (Sept., 2009) 321--332,
  [\href{http://arxiv.org/abs/0902.2013}{{\tt arXiv:0902.2013}}].

\bibitem{desjacques/seljak/iliev:2009}
V.~{Desjacques}, U.~{Seljak}, and I.~T. {Iliev}, {\it {Scale-dependent bias
  induced by local non-Gaussianity: a comparison to N-body simulations}},  {\em
  Mon. Not. R. Astron. Soc.} {\bf 396} (June, 2009) 85--96,
  [\href{http://arxiv.org/abs/0811.2748}{{\tt arXiv:0811.2748}}].

\bibitem{2010MNRAS.402..191P}
A.~{Pillepich}, C.~{Porciani}, and O.~{Hahn}, {\it {Halo mass function and
  scale-dependent bias from N-body simulations with non-Gaussian initial
  conditions}},  {\em \mnras} {\bf 402} (Feb., 2010) 191--206,
  [\href{http://arxiv.org/abs/0811.4176}{{\tt arXiv:0811.4176}}].

\bibitem{baldauf/etal:2015}
T.~{Baldauf}, U.~{Seljak}, L.~{Senatore}, and M.~{Zaldarriaga}, {\it {Linear
  response to long wavelength fluctuations using curvature simulations}},  {\em
  \jcap} {\bf 9} (Sept., 2016) 007,
  [\href{http://arxiv.org/abs/1511.01465}{{\tt arXiv:1511.01465}}].

\bibitem{2017MNRAS.468.3277B}
M.~{Biagetti}, T.~{Lazeyras}, T.~{Baldauf}, V.~{Desjacques}, and F.~{Schmidt},
  {\it {Verifying the consistency relation for the scale-dependent bias from
  local primordial non-Gaussianity}},  {\em \mnras} {\bf 468} (Jul, 2017)
  3277--3288, [\href{http://arxiv.org/abs/1611.04901}{{\tt arXiv:1611.04901}}].

\bibitem{2020JCAP...12..013B}
A.~{Barreira}, G.~{Cabass}, F.~{Schmidt}, A.~{Pillepich}, and D.~{Nelson}, {\it
  {Galaxy bias and primordial non-Gaussianity: insights from galaxy formation
  simulations with IllustrisTNG}},  {\em \jcap} {\bf 2020} (Dec., 2020) 013,
  [\href{http://arxiv.org/abs/2006.09368}{{\tt arXiv:2006.09368}}].

\bibitem{2022JCAP...01..033B}
A.~{Barreira}, {\it {Predictions for local PNG bias in the galaxy power
  spectrum and bispectrum and the consequences for f $_{NL}$ constraints}},
  {\em \jcap} {\bf 2022} (Jan., 2022) 033,
  [\href{http://arxiv.org/abs/2107.06887}{{\tt arXiv:2107.06887}}].

\bibitem{2022JCAP...04..057B}
A.~{Barreira}, {\it {The local PNG bias of neutral Hydrogen, H$_{I}$}},  {\em
  \jcap} {\bf 2022} (Apr., 2022) 057,
  [\href{http://arxiv.org/abs/2112.03253}{{\tt arXiv:2112.03253}}].

\bibitem{Pillepich:2017jle}
A.~Pillepich et~al., {\it {Simulating Galaxy Formation with the IllustrisTNG
  Model}},  {\em Mon. Not. Roy. Astron. Soc.} {\bf 473} (2018), no.~3
  4077--4106, [\href{http://arxiv.org/abs/1703.02970}{{\tt arXiv:1703.02970}}].

\bibitem{2019MNRAS.488.2079B}
A.~{Barreira}, D.~{Nelson}, A.~{Pillepich}, V.~{Springel}, F.~{Schmidt},
  R.~{Pakmor}, L.~{Hernquist}, and M.~{Vogelsberger}, {\it {Separate Universe
  simulations with IllustrisTNG: baryonic effects on power spectrum responses
  and higher-order statistics}},  {\em \mnras} {\bf 488} (Sept., 2019)
  2079--2092, [\href{http://arxiv.org/abs/1904.02070}{{\tt arXiv:1904.02070}}].

\bibitem{2022arXiv220505673B}
A.~{Barreira}, {\it {Can we actually constrain $f_{\rm NL}$ using the
  scale-dependent bias effect? An illustration of the impact of galaxy bias
  uncertainties using the BOSS DR12 galaxy power spectrum}},  {\em arXiv
  e-prints} (May, 2022) arXiv:2205.05673,
  [\href{http://arxiv.org/abs/2205.05673}{{\tt arXiv:2205.05673}}].

\bibitem{2020JCAP...12..031B}
A.~{Barreira}, {\it {On the impact of galaxy bias uncertainties on primordial
  non-Gaussianity constraints}},  {\em \jcap} {\bf 2020} (Dec., 2020) 031,
  [\href{http://arxiv.org/abs/2009.06622}{{\tt arXiv:2009.06622}}].

\bibitem{2021JCAP...05..015M}
A.~{Moradinezhad Dizgah}, M.~{Biagetti}, E.~{Sefusatti}, V.~{Desjacques}, and
  J.~{Nore{\~n}a}, {\it {Primordial non-Gaussianity from biased tracers:
  likelihood analysis of real-space power spectrum and bispectrum}},  {\em
  \jcap} {\bf 2021} (May, 2021) 015,
  [\href{http://arxiv.org/abs/2010.14523}{{\tt arXiv:2010.14523}}].

\bibitem{Sheth:2004}
R.~K. Sheth and G.~Tormen, {\it {On the environmental dependence of halo
  formation}},  {\em Mon. Not. Roy. Astron. Soc.} {\bf 350} (2004) 1385,
  [\href{http://arxiv.org/abs/astro-ph/0402237}{{\tt astro-ph/0402237}}].

\bibitem{gao:2005}
L.~Gao, V.~Springel, and S.~D.~M. White, {\it {The Age dependence of halo
  clustering}},  {\em Mon. Not. Roy. Astron. Soc.} {\bf 363} (2005) L66--L70,
  [\href{http://arxiv.org/abs/astro-ph/0506510}{{\tt astro-ph/0506510}}].

\bibitem{Gao:2006}
L.~Gao and S.~D.~M. White, {\it {Assembly bias in the clustering of dark matter
  haloes}},  {\em Mon. Not. Roy. Astron. Soc.} {\bf 377} (2007) L5--L9,
  [\href{http://arxiv.org/abs/astro-ph/0611921}{{\tt astro-ph/0611921}}].

\bibitem{Wechsler:2005}
R.~H. Wechsler, A.~R. Zentner, J.~S. Bullock, and A.~V. Kravtsov, {\it {The
  dependence of halo clustering on halo formation history, concentration, and
  occupation}},  {\em Astrophys. J.} {\bf 652} (2006) 71--84,
  [\href{http://arxiv.org/abs/astro-ph/0512416}{{\tt astro-ph/0512416}}].

\bibitem{Jing:2006}
Y.~P. Jing, Y.~Suto, and H.~J. Mo, {\it {The dependence of dark halo clustering
  on the formation epoch and the concentration parameter}},  {\em Astrophys.
  J.} {\bf 657} (2007) 664--668,
  [\href{http://arxiv.org/abs/astro-ph/0610099}{{\tt astro-ph/0610099}}].

\bibitem{Croton:2006}
D.~J. Croton, L.~Gao, and S.~D.~M. White, {\it {Halo assembly bias and its
  effects on galaxy clustering}},  {\em Mon. Not. Roy. Astron. Soc.} {\bf 374}
  (2007) 1303--1309, [\href{http://arxiv.org/abs/astro-ph/0605636}{{\tt
  astro-ph/0605636}}].

\bibitem{Angulo:2007}
R.~E. Angulo, C.~M. Baugh, and C.~G. Lacey, {\it {The assembly bias of dark
  matter haloes to higher orders}},  {\em Mon. Not. Roy. Astron. Soc.} {\bf
  387} (2008) 921, [\href{http://arxiv.org/abs/0712.2280}{{\tt
  arXiv:0712.2280}}].

\bibitem{Dalal:2008}
N.~Dalal, M.~White, J.~R. Bond, and A.~Shirokov, {\it {Halo Assembly Bias in
  Hierarchical Structure Formation}},  {\em Astrophys. J.} {\bf 687} (2008)
  12--21, [\href{http://arxiv.org/abs/0803.3453}{{\tt arXiv:0803.3453}}].

\bibitem{Faltenbacher:2009}
A.~Faltenbacher and S.~D.~M. White, {\it {Assembly bias and the dynamical
  structure of dark matter halos}},  {\em Astrophys. J.} {\bf 708} (2010)
  469--473, [\href{http://arxiv.org/abs/0909.4302}{{\tt arXiv:0909.4302}}].

\bibitem{Lacerna:2012}
I.~Lacerna and N.~Padilla, {\it {The nature of assembly bias - II. Halo spin}},
   {\em Mon. Not. Roy. Astron. Soc.} {\bf 426} (2012) 26,
  [\href{http://arxiv.org/abs/1207.4476}{{\tt arXiv:1207.4476}}].

\bibitem{Sunayama:2015}
T.~Sunayama, A.~P. Hearin, N.~Padmanabhan, and A.~Leauthaud, {\it {The
  Scale-Dependence of Halo Assembly Bias}},  {\em Mon. Not. Roy. Astron. Soc.}
  {\bf 458} (2016), no.~2 1510--1516,
  [\href{http://arxiv.org/abs/1509.06417}{{\tt arXiv:1509.06417}}].

\bibitem{Paranjape:2016}
A.~Paranjape and N.~Padmanabhan, {\it {Halo assembly bias from Separate
  Universe simulations}},  {\em Mon. Not. Roy. Astron. Soc.} {\bf 468} (2017),
  no.~3 2984--2999, [\href{http://arxiv.org/abs/1612.02833}{{\tt
  arXiv:1612.02833}}].

\bibitem{Lazeyras:2016}
T.~Lazeyras, M.~Musso, and F.~Schmidt, {\it {Large-scale assembly bias of dark
  matter halos}},  {\em JCAP} {\bf 1703} (2017), no.~03 059,
  [\href{http://arxiv.org/abs/1612.04360}{{\tt arXiv:1612.04360}}].

\bibitem{Salcedo:2017}
A.~N. Salcedo, A.~H. Maller, A.~A. Berlind, M.~Sinha, C.~K. McBride, P.~S.
  Behroozi, R.~H. Wechsler, and D.~H. Weinberg, {\it {Spatial clustering of
  dark matter haloes: secondary bias, neighbour bias, and the influence of
  massive neighbours on halo properties}},  {\em Mon. Not. Roy. Astron. Soc.}
  {\bf 475} (2018), no.~4 4411--4423,
  [\href{http://arxiv.org/abs/1708.08451}{{\tt arXiv:1708.08451}}].

\bibitem{Mao:2017}
Y.-Y. Mao, A.~R. Zentner, and R.~H. Wechsler, {\it {Beyond Assembly Bias:
  Exploring Secondary Halo Biases for Cluster-size Haloes}},  {\em Mon. Not.
  Roy. Astron. Soc.} {\bf 474} (2018), no.~4 5143--5157,
  [\href{http://arxiv.org/abs/1705.03888}{{\tt arXiv:1705.03888}}]. [Erratum:
  Mon.Not.Roy.Astron.Soc. 481, 3167 (2018)].

\bibitem{Chue:2018}
C.~Y.~R. Chue, N.~Dalal, and M.~White, {\it {Some assembly required: assembly
  bias in massive dark matter halos}},  {\em JCAP} {\bf 10} (2018) 012,
  [\href{http://arxiv.org/abs/1804.04055}{{\tt arXiv:1804.04055}}].

\bibitem{Sato-Polito:2018}
G.~Sato-Polito, A.~D. Montero-Dorta, L.~R. Abramo, F.~Prada, and A.~Klypin,
  {\it {The dependence of halo bias on age, concentration and spin}},  {\em
  Mon. Not. Roy. Astron. Soc.} {\bf 487} (2019), no.~2 1570--1579,
  [\href{http://arxiv.org/abs/1810.02375}{{\tt arXiv:1810.02375}}].

\bibitem{Paco_18}
F.~{Villaescusa-Navarro}, S.~{Genel}, E.~{Castorina}, A.~{Obuljen}, D.~N.
  {Spergel}, L.~{Hernquist}, D.~{Nelson}, I.~P. {Carucci}, A.~{Pillepich},
  F.~{Marinacci}, B.~{Diemer}, M.~{Vogelsberger}, R.~{Weinberger}, and
  R.~{Pakmor}, {\it {Ingredients for 21 cm Intensity Mapping}},  {\em \apj}
  {\bf 866} (Oct., 2018) 135, [\href{http://arxiv.org/abs/1804.09180}{{\tt
  arXiv:1804.09180}}].

\bibitem{Lazeyras:2020suj}
T.~Lazeyras, F.~Villaescusa-Navarro, and M.~Viel, {\it {The impact of massive
  neutrinos on halo assembly bias}},  {\em JCAP} {\bf 03} (2021) 022,
  [\href{http://arxiv.org/abs/2008.12265}{{\tt arXiv:2008.12265}}].

\bibitem{Contreras:2021oxg}
S.~Contreras, J.~Chaves-Montero, M.~Zennaro, and R.~E. Angulo, {\it {The
  cosmological dependence of halo and galaxy assembly bias}},
  \href{http://arxiv.org/abs/2105.05854}{{\tt arXiv:2105.05854}}.

\bibitem{Lazeyras:2021dar}
T.~Lazeyras, A.~Barreira, and F.~Schmidt, {\it {Assembly bias in quadratic bias
  parameters of dark matter halos from forward modeling}},  {\em JCAP} {\bf 10}
  (2021) 063, [\href{http://arxiv.org/abs/2106.14713}{{\tt arXiv:2106.14713}}].

\bibitem{Reid:2010}
B.~A. {Reid}, L.~{Verde}, K.~{Dolag}, S.~{Matarrese}, and L.~{Moscardini}, {\it
  {Non-Gaussian halo assembly bias}},  {\em \jcap} {\bf 7} (July, 2010) 013,
  [\href{http://arxiv.org/abs/1004.1637}{{\tt arXiv:1004.1637}}].

\bibitem{Springel:2005}
V.~Springel, {\it {The Cosmological simulation code GADGET-2}},  {\em
  Mon.Not.Roy.Astron.Soc.} {\bf 364} (2005) 1105--1134,
  [\href{http://arxiv.org/abs/astro-ph/0505010}{{\tt astro-ph/0505010}}].

\bibitem{sirko:2005}
E.~Sirko, {\it {Initial conditions to cosmological N-body simulations, or how
  to run an ensemble of simulations}},  {\em \apj} {\bf 634} (2005) 728--743,
  [\href{http://arxiv.org/abs/astro-ph/0503106}{{\tt astro-ph/0503106}}].

\bibitem{EUCLID:2011zbd}
{\bf EUCLID} Collaboration, R.~Laureijs et~al., {\it {Euclid Definition Study
  Report}},  \href{http://arxiv.org/abs/1110.3193}{{\tt arXiv:1110.3193}}.

\bibitem{Gill:2004}
S.~P. Gill, A.~Knebe, and B.~K. Gibson, {\it {The Evolution substructure 1: A
  New identification method}},  {\em Mon.Not.Roy.Astron.Soc.} {\bf 351} (2004)
  399, [\href{http://arxiv.org/abs/astro-ph/0404258}{{\tt astro-ph/0404258}}].

\bibitem{Knollmann:2009}
S.~R. Knollmann and A.~Knebe, {\it {Ahf: Amiga's Halo Finder}},  {\em
  Astrophys.J.Suppl.} {\bf 182} (2009) 608--624,
  [\href{http://arxiv.org/abs/0904.3662}{{\tt arXiv:0904.3662}}].

\bibitem{Scoccimarro:1997gr}
R.~Scoccimarro, {\it {Transients from initial conditions: a perturbative
  analysis}},  {\em Mon. Not. Roy. Astron. Soc.} {\bf 299} (1998) 1097,
  [\href{http://arxiv.org/abs/astro-ph/9711187}{{\tt astro-ph/9711187}}].

\bibitem{Crocce:2006ve}
M.~Crocce, S.~Pueblas, and R.~Scoccimarro, {\it {Transients from Initial
  Conditions in Cosmological Simulations}},  {\em Mon. Not. Roy. Astron. Soc.}
  {\bf 373} (2006) 369--381, [\href{http://arxiv.org/abs/astro-ph/0606505}{{\tt
  astro-ph/0606505}}].

\bibitem{Biagetti:2020skr}
M.~Biagetti, A.~Cole, and G.~Shiu, {\it {The Persistence of Large Scale
  Structures I: Primordial non-Gaussianity}},  {\em JCAP} {\bf 04} (2021) 061,
  [\href{http://arxiv.org/abs/2009.04819}{{\tt arXiv:2009.04819}}].

\bibitem{2010MNRAS.401..791S}
V.~{Springel}, {\it {E pur si muove: Galilean-invariant cosmological
  hydrodynamical simulations on a moving mesh}},  {\em \mnras} {\bf 401} (Jan,
  2010) 791--851, [\href{http://arxiv.org/abs/0901.4107}{{\tt
  arXiv:0901.4107}}].

\bibitem{2020ApJS..248...32W}
R.~{Weinberger}, V.~{Springel}, and R.~{Pakmor}, {\it {The AREPO Public Code
  Release}},  {\em \apjs} {\bf 248} (June, 2020) 32,
  [\href{http://arxiv.org/abs/1909.04667}{{\tt arXiv:1909.04667}}].

\bibitem{2015ascl.soft02003S}
V.~{Springel}, ``{N-GenIC: Cosmological structure initial conditions}.''
  Astrophysics Source Code Library, Feb., 2015.

\bibitem{Lazeyras:2017}
T.~Lazeyras and F.~Schmidt, {\it {Beyond LIMD bias: a measurement of the
  complete set of third-order halo bias parameters}},  {\em JCAP} {\bf 1809}
  (2018), no.~09 008, [\href{http://arxiv.org/abs/1712.07531}{{\tt
  arXiv:1712.07531}}].

\bibitem{Navarro:1996}
J.~F. Navarro, C.~S. Frenk, and S.~D.~M. White, {\it {A Universal density
  profile from hierarchical clustering}},  {\em Astrophys. J.} {\bf 490} (1997)
  493--508, [\href{http://arxiv.org/abs/astro-ph/9611107}{{\tt
  astro-ph/9611107}}].

\bibitem{Prada:2011}
F.~Prada, A.~A. Klypin, A.~J. Cuesta, J.~E. Betancort-Rijo, and J.~Primack,
  {\it {Halo concentrations in the standard LCDM cosmology}},  {\em Mon. Not.
  Roy. Astron. Soc.} {\bf 423} (2012) 3018--3030,
  [\href{http://arxiv.org/abs/1104.5130}{{\tt arXiv:1104.5130}}].

\bibitem{Bullock:2000}
J.~S. Bullock, A.~Dekel, T.~S. Kolatt, A.~V. Kravtsov, A.~A. Klypin,
  C.~Porciani, and J.~R. Primack, {\it {A Universal angular momentum profile
  for galactic halos}},  {\em Astrophys. J.} {\bf 555} (2001) 240--257,
  [\href{http://arxiv.org/abs/astro-ph/0011001}{{\tt astro-ph/0011001}}].

\bibitem{Bullock:1999he}
J.~S. Bullock, T.~S. Kolatt, Y.~Sigad, R.~S. Somerville, A.~V. Kravtsov, A.~A.
  Klypin, J.~R. Primack, and A.~Dekel, {\it {Profiles of dark haloes.
  Evolution, scatter, and environment}},  {\em Mon. Not. Roy. Astron. Soc.}
  {\bf 321} (2001) 559--575, [\href{http://arxiv.org/abs/astro-ph/9908159}{{\tt
  astro-ph/9908159}}].

\bibitem{Wechsler:2001}
R.~H. Wechsler, J.~S. Bullock, J.~R. Primack, A.~V. Kravtsov, and A.~Dekel,
  {\it {Concentrations of dark halos from their assembly histories}},  {\em
  Astrophys. J.} {\bf 568} (2002) 52--70,
  [\href{http://arxiv.org/abs/astro-ph/0108151}{{\tt astro-ph/0108151}}].

\bibitem{Voivodic:2020bec}
R.~Voivodic and A.~Barreira, {\it {Responses of Halo Occupation Distributions:
  a new ingredient in the halo model \& the impact on galaxy bias}},  {\em
  JCAP} {\bf 05} (2021) 069, [\href{http://arxiv.org/abs/2012.04637}{{\tt
  arXiv:2012.04637}}].

\bibitem{Barreira:2021ukk}
A.~Barreira, T.~Lazeyras, and F.~Schmidt, {\it {Galaxy bias from forward
  models: linear and second-order bias of IllustrisTNG galaxies}},
  \href{http://arxiv.org/abs/2105.02876}{{\tt arXiv:2105.02876}}.

\bibitem{2022arXiv220712398O}
A.~{Obuljen}, M.~{Simonovi{\'c}}, A.~{Schneider}, and R.~{Feldmann}, {\it
  {Modeling HI at the field level}},  {\em arXiv e-prints} (July, 2022)
  arXiv:2207.12398, [\href{http://arxiv.org/abs/2207.12398}{{\tt
  arXiv:2207.12398}}].

\bibitem{2022PhRvD.105d3517P}
O.~H.~E. {Philcox} and M.~M. {Ivanov}, {\it {BOSS DR12 full-shape cosmology:
  {\ensuremath{\Lambda}} CDM constraints from the large-scale galaxy power
  spectrum and bispectrum monopole}},  {\em \prd} {\bf 105} (Feb., 2022)
  043517, [\href{http://arxiv.org/abs/2112.04515}{{\tt arXiv:2112.04515}}].

\bibitem{2011ApJ...728..126W}
M.~{White} and {et al}, {\it {The Clustering of Massive Galaxies at z
  \raisebox{-0.5ex}\textasciitilde 0.5 from the First Semester of BOSS Data}},
  {\em \apj} {\bf 728} (Feb., 2011) 126,
  [\href{http://arxiv.org/abs/1010.4915}{{\tt arXiv:1010.4915}}].

\bibitem{2013MNRAS.429...98P}
J.~K. {Parejko} and {et al}, {\it {The clustering of galaxies in the SDSS-III
  Baryon Oscillation Spectroscopic Survey: the low-redshift sample}},  {\em
  \mnras} {\bf 429} (Feb., 2013) 98--112,
  [\href{http://arxiv.org/abs/1211.3976}{{\tt arXiv:1211.3976}}].

\bibitem{2016MNRAS.455.1553R}
B.~{Reid} and {et al}, {\it {SDSS-III Baryon Oscillation Spectroscopic Survey
  Data Release 12: galaxy target selection and large-scale structure
  catalogues}},  {\em \mnras} {\bf 455} (Jan., 2016) 1553--1573,
  [\href{http://arxiv.org/abs/1509.06529}{{\tt arXiv:1509.06529}}].

\bibitem{2017ApJ...840..104S}
H.~{Shan}, J.-P. {Kneib}, R.~{Li}, J.~{Comparat}, T.~{Erben}, M.~{Makler},
  B.~{Moraes}, L.~{Van Waerbeke}, J.~E. {Taylor}, A.~{Charbonnier}, and
  M.~E.~S. {Pereira}, {\it {The Mass-Concentration Relation and the
  Stellar-to-halo Mass Ratio in the CFHT Stripe 82 Survey}},  {\em \apj} {\bf
  840} (May, 2017) 104, [\href{http://arxiv.org/abs/1502.00313}{{\tt
  arXiv:1502.00313}}].

\bibitem{2021PhRvD.103j3504P}
O.~H.~E. {Philcox}, {\it {Cosmology without window functions: Quadratic
  estimators for the galaxy power spectrum}},  {\em \prd} {\bf 103} (May, 2021)
  103504, [\href{http://arxiv.org/abs/2012.09389}{{\tt arXiv:2012.09389}}].

\bibitem{2016MNRAS.456.4156K}
F.-S. {Kitaura} and {et al}, {\it {The clustering of galaxies in the SDSS-III
  Baryon Oscillation Spectroscopic Survey: mock galaxy catalogues for the BOSS
  Final Data Release}},  {\em \mnras} {\bf 456} (Mar., 2016) 4156--4173,
  [\href{http://arxiv.org/abs/1509.06400}{{\tt arXiv:1509.06400}}].

\bibitem{2016MNRAS.460.1173R}
S.~A. {Rodr{\'\i}guez-Torres} and {et al}, {\it {The clustering of galaxies in
  the SDSS-III Baryon Oscillation Spectroscopic Survey: modelling the
  clustering and halo occupation distribution of BOSS CMASS galaxies in the
  Final Data Release}},  {\em \mnras} {\bf 460} (Aug., 2016) 1173--1187,
  [\href{http://arxiv.org/abs/1509.06404}{{\tt arXiv:1509.06404}}].

\bibitem{Maccio:2006wpz}
A.~V. Macci\`o, A.~A. Dutton, F.~C. van~den Bosch, B.~Moore, D.~Potter, and
  J.~Stadel, {\it {Concentration, Spin and Shape of Dark Matter Haloes: Scatter
  and the Dependence on Mass and Environment}},  {\em Mon. Not. Roy. Astron.
  Soc.} {\bf 378} (2007) 55--71,
  [\href{http://arxiv.org/abs/astro-ph/0608157}{{\tt astro-ph/0608157}}].

\end{thebibliography}\endgroup
\end{document}